\documentstyle{ar}
\begin{document}
\input{psfig.sty}
\input{epsf.sty}

\title{Prospects for Spin Physics at RHIC}  
\markboth{Bunce et al}{Prospects for Spin Physics at RHIC}  
\author{Gerry Bunce\affiliation{Brookhaven National Laboratory, Upton,
New York 11973-5000\\
and\\
RIKEN BNL Research Center,
Brookhaven National Laboratory, Upton,
New York 11973-5000\\
{\tt email:~bunce@bnl.gov}}
Naohito Saito\affiliation{
RIKEN (The Institute of Physical and Chemical Research),
Wako, Saitama 351-0198, Japan \\
and \\
RIKEN BNL Research Center,
Brookhaven National Laboratory, Upton,
New York 11973-5000\\
{\tt email:~saito@bnl.gov}}
Jacques Soffer\affiliation{Centre de Physique Th\'{e}orique--CNRS--
Luminy, Case 907, F-13288 Marseille Cedex 9, France\\
{\tt email:~Jacques.Soffer@cpt.univ-mrs.fr}}
Werner Vogelsang\affiliation{C.N.\ Yang Institute for Theoretical
Physics, State University of New York at Stony Brook, Stony Brook,
New York 11794-3840\\
and \\
RIKEN BNL Research Center,
Brookhaven National Laboratory, Upton,
New York 11973-5000\\
(present address)\\
{\tt email:~wvogelsang@bnl.gov}}}
\begin{abstract}
        Colliding beams of 70\% polarized protons at up to $\sqrt{s}$=500~GeV,
with high luminosity, L=2$\times$10$^{{\rm 32}}$~cm$^{-2}$sec$^{-1}$, will
represent a new and unique laboratory for studying the proton. RHIC-Spin
will be the first polarized-proton collider and will be capable of copious
production of jets, directly produced photons, and $W$ and $Z$ bosons. 
Features will include
direct and precise measurements of the polarization of the gluons and of
$\bar{u}$, $\bar{d}$, $u$, and $d$ quarks in a polarized proton.
Parity violation searches for physics beyond the standard model will
be competitive with unpolarized searches at the Fermilab Tevatron. 
Transverse spin will explore transversity for the first time, as well as quark-gluon
correlations in the proton.  Spin dependence of the total cross section and in
the Coulomb nuclear interference region will be measured at collider energies for the first time.
These qualitatively new measurements can be expected 
to deepen our understanding of the structure of matter and of the strong interaction.

\end{abstract}
\maketitle

\section{INTRODUCTION} 
Spin is a powerful and elegant tool in physics.  
One of the most exciting aspects of physics 
is a search for the unexpected, the nonintuitive, in nature.  
Intrinsic spin itself violates our intuition, in that an elementary particle 
such as an electron can both be pointlike 
and have a perpetual angular momentum.  
We find at this time an apparent violation of our intuition in the proton.  
We understand the proton as being composed of quarks, gluons, and 
antiquarks, and we expect the proton spin to be 
carried dominantly by its three valence quarks.  
Instead, through the 1980s and 1990s, deep inelastic scattering (DIS) experiments 
of polarized electrons and muons from polarized nucleons 
have shown that on average only about 1/4 to 1/3 of the proton spin is 
carried by the quarks and antiquarks in the proton~\cite{dsigma}.  
Therefore, the spin of the proton appears to be mainly carried
by the gluons and orbital angular momentum! This surprising and 
counterintuitive result indicates that the proton, and particularly 
its spin structure,  is much more interesting than we had 
thought.

Spin can be used as an elegant tool to search for the unexpected. 
If an experiment is found to depend on the spin direction, it can 
violate a deep expectation that physics should be 
symmetric with respect to that axis. An example is mirror symmetry, that 
physics should not depend on left- or right-handedness.  The violation of 
parity by the weak interaction was the surprise that led to the present 
electroweak model with the purely left-handed charged weak vector bosons 
$W^{\pm}$. At the Relativistic Heavy Ion Collider (RHIC) 
at Brookhaven National Laboratory, the $W^+$ and $W^-$ will be produced 
by colliding beams of protons spinning alternately left- and right-handed.  
The expected maximum violation of parity will allow unique and precise 
measurements of the spin direction 
of the quarks and antiquarks in the proton that form the $W$ bosons, 
identified by quark flavor, $u$, $\bar{u}$, $d$, and $\bar{d}$. 
A dependence on handedness in the production of 
jets at RHIC beyond the contribution from $W$ and $Z$ 
would directly signal new physics, possibly coming from quark substructure 
at a scale above the weak scale.

Physics is also a search for unexpected order in nature. Large spin effects 
necessarily imply coherence and order. If the gluons in a proton are 
found to be dominantly spinning in the same direction, as discussed 
widely in the context of the smallness of the quark spin 
contribution~(reviewed in~\cite{bass99}), there would need to be a simple underlying 
physical mechanism that creates this order. 
At RHIC, dedicated experiments will measure the direction of the gluon 
spin in the proton for the first time---an exciting prospect, since there 
are hints that the gluon polarization may be substantial. 

The RHIC at Brookhaven has begun a program of colliding beams of gold ions at 
100~GeV per nucleon in the spring of 2000.  The following year, the first 
physics run colliding beams of polarized protons is expected.  RHIC-Spin 
will be the first polarized proton-proton 
collider. It will reach an energy 
%
and luminosity at which the collisions can clearly be interpreted as 
collisions of 
polarized quarks and gluons, and it will be capable of copious production of jets and 
directly produced photons, 
as well as $W$ and $Z$ bosons. Quantum chromodynamics (QCD) 
makes definite predictions
for the hard spin interactions of quarks and gluons, which implies
that RHIC 
will enable us to test a sector of QCD that so far has been little
explored. The  polarized quark and gluon probes at RHIC complement the 
beautiful work done using polarized lepton probes to study proton spin 
structure. These strong interaction probes will be sensitive to 
the gluon polarization in jet and direct photon production and will allow 
 quark spin-flavor separation in $W^{\pm}$ production. RHIC-Spin will also 
represent the highest energy for proton-proton collisions at accelerators, 
and unpolarized $W^{\pm}$ production will be used to precisely measure 
the flavor asymmetry of the antiquark sea.

At the Polarized Collider Workshop at Penn State University 
in 1990~\cite{collins91}, the exploration of the spin of the proton was 
a major 
focus for the physics of polarized proton collisions at RHIC.  The RHIC 
Spin Collaboration was formed the following year, consisting of experimenters, 
theorists, and accelerator physicists~\cite{Bunce91}. Since 1993, the two large heavy ion detectors at 
RHIC, {\sc Star} and {\sc Phenix}, have considered spin as 
a major program and include additional apparatus specifically for spin physics. 
In addition, the {\sc pp2pp} experiment at RHIC, studying small-angle elastic scattering, will also feature spin.  
The present article presents the anticipated 
physics of the RHIC spin program as developed by the RHIC Spin Collaboration 
and by the {\sc Star}, {\sc Phenix}, and {\sc pp2pp} Collaborations.

\begin{figure}[h] 
  \centerline{\psfig{figure=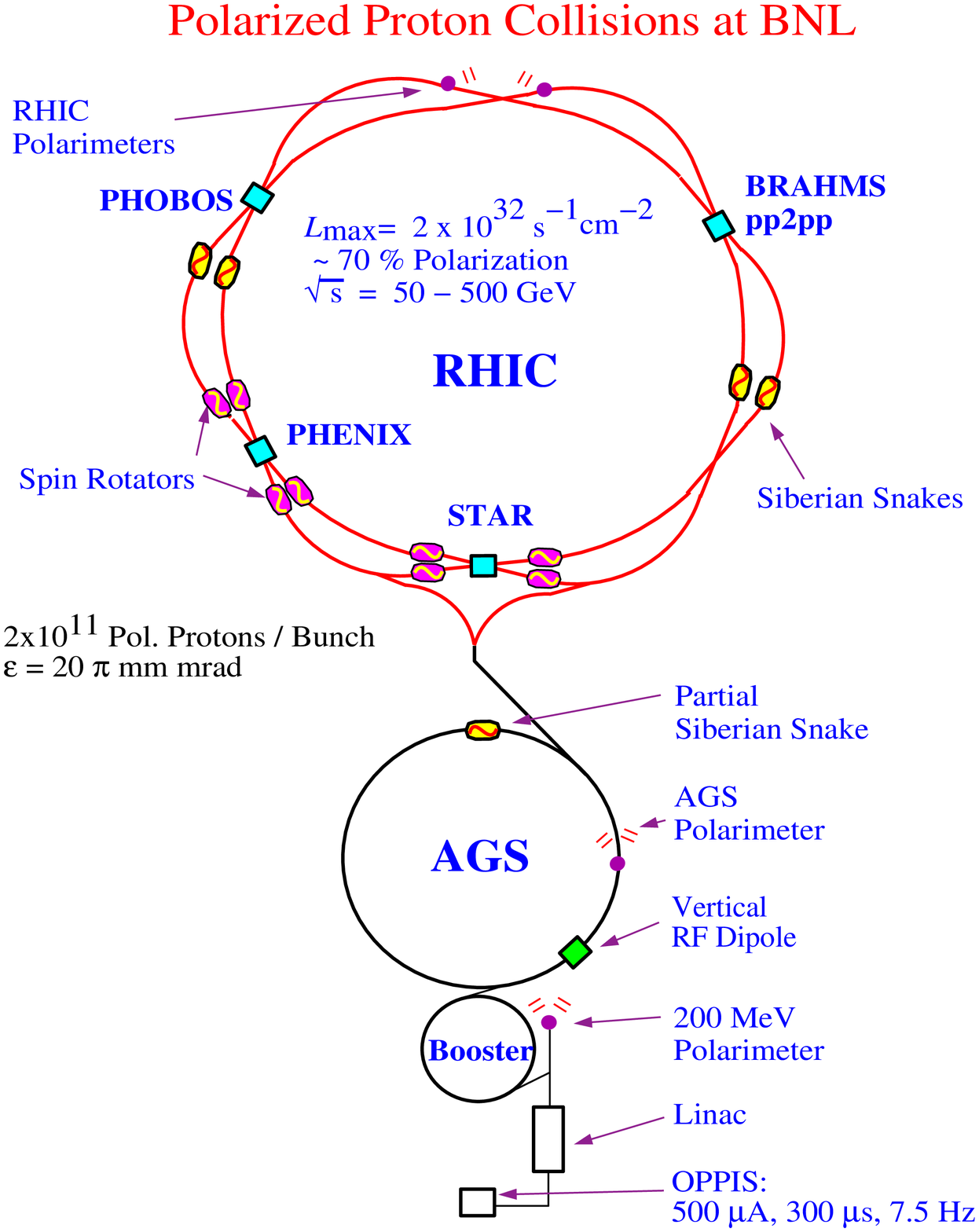,width=12cm,height=15cm,angle=0}}
\caption{\small Schematic layout of the RHIC 
accelerator complex. Only relevant devices for polarized $pp$ collisions
are shown. }
\label{F:RHICSPIN}
\end{figure}

The RHIC spin accelerator complex is illustrated in Figure~1.  An intense 
polarized $H^-$ source feeds a chain of accelerators. 
Individual bunches of $2\times 10^{11}$ protons with 70\% polarization are 
transferred from the Alternating Gradient Synchrotron (AGS) to the RHIC rings at 22~GeV. This is repeated 
120 times for each ring at RHIC. The polarized protons are then accelerated
to up to 250~GeV in each ring for collisions at each of 6 intersection 
regions. 
With a $\beta ^* = 1$--meter focus at {\sc Star} and {\sc Phenix}, luminosity 
will be ${\cal L}=2\times 10^{32}~$cm$^{-2}$s$^{-1}$, for the highest 
RHIC energy of $\sqrt s = 500$~GeV. Experimental sensitivities given in 
this article are based on 800~pb$^{-1}$ for 
$\sqrt{s} = 500$~GeV and $320$~pb$^{-1}$ 
for  $\sqrt{s} = 200$~GeV.  This corresponds to runs of $4\times 10^6$~s at 
full luminosity, about four months of running with 40\% efficiency, at each energy.
We expect the data to be collected 
over three to four years, since RHIC is shared between heavy-ion and 
polarized-proton collisions. 
The expected sensitivities will be excellent due to
the high luminosity for proton-proton collisions. 
For comparison, we note that the $\bar{p}p$ Tevatron at Fermilab 
has run for a total of $\sim$ 130~pb$^{-1}$ as of 1999.    

It is difficult to maintain the proton polarization through acceleration because of 
its large anomalous magnetic moment: 
the proton spin readily responds to focusing and error magnetic
fields in the rings, and spin resonances are encountered frequently, for 
example at every 500 MeV of acceleration in the AGS. 
The methods that are used to avoid depolarization in acceleration 
are very elegant, and the acceleration 
of polarized protons to 250 GeV will be breaking new ground in accelerator 
physics. The key device is a string of dipole magnets that rotate the 
proton spin 180$^{\circ}$ around 
a selected axis in the horizontal plane each time the beam 
passes~\cite{derbenev}.  Each two passes in effect cancel 
the cumulative tilt of the spin resulting from horizontal magnetic fields, thus eliminating 
the major spin resonances at RHIC. 
There will be four ``Siberian Snakes" at RHIC, two in each ring. The name 
refers to the home institution of the inventors (Novosibirsk) and to the motion of the beam passing through.  In this article, we 
do not discuss the accelerator physics work leading to the RHIC spin plan~\cite{alekseev}, but, 
as for any spin experiment, past or future, 
there is a very tight, necessary, and refreshing coupling between the 
polarization technology and the physics.

For two Siberian Snakes in each ring, the stable spin direction in RHIC 
will be vertical. Therefore, transverse 
spin physics will be available to all the experiments.  
For {\sc Star} and {\sc Phenix}, special strings of dipole magnets 
will be used to rotate the spin to longitudinal at their intersection 
regions.  Longitudinal spin is necessary 
to study gluon polarization and parity-violating physics.  A recent 
plan~\cite{lerach} is to initially use one 
Siberian Snake in each ring, which allows the construction and installation 
of the Snakes and Rotators to be staged.  
With a single Snake in a ring, the stable spin direction is in the horizontal 
plane.  If the beam is inserted into RHIC, 
and the Snake is then turned on adiabatically, the spin will follow from 
vertical to horizontal.  At energies roughly 2~GeV apart, 
it will be possible to have longitudinal polarization at all six 
intersection regions, up to a beam energy of 100 GeV.  
One Snake is already installed in RHIC at this time, and a 
second Snake will be completed in summer 2000.  
Therefore, the RHIC-Spin program will be ready for its commissioning 
in summer 2000  and ready for the first spin physics 
run with longitudinal polarization at $\sqrt s = 200$~GeV in 2001.

\newcommand\sss{\scriptscriptstyle}
\newcommand\gs{g_{\sss S}}
\newcommand\as{\alpha_{\sss S}}         
\newcommand\ep{\epsilon}
\newcommand\Th{\theta}
\newcommand\epb{\overline{\epsilon}}
\newcommand\mug{\mu_\gamma}
\newcommand\mue{\mu_e}
\newcommand\muf{\mu_{\sss F}}
\newcommand\mufp{\mu_{\sss F}^\prime}
\newcommand\mufs{\mu_{\sss F}^{\prime\prime}}
\newcommand\mur{\mu_{\sss R}}
\newcommand\murp{\mu_{\sss R}^\prime}
\newcommand\murs{\mu_{\sss R}^{\prime\prime}}
\newcommand\muh{\mu_{\sss H}}
\newcommand\muhp{\mu_{\sss H}^\prime}
\newcommand\muhs{\mu_{\sss H}^{\prime\prime}}
\newcommand\muo{\mu_0}
\newcommand\MSB{{\rm \overline{MS}}}
\newcommand\DIG{{\rm DIS}_\gamma}
\newcommand\CA{C_{\sss A}}
\newcommand\DA{D_{\sss A}}
\newcommand\CF{C_{\sss F}}
\newcommand\TF{T_{\sss F}}
\newcommand\pt{p_{\sss T}}
\newcommand\kt{k_{\sss T}}
\newcommand\ptg{p_{{\sss T}\gamma}}
\newcommand\xtg{x_{\sss T}^\gamma}
\newcommand\etag{\eta_\gamma}
\newcommand\phig{\phi_\gamma}
\newcommand\ptj{p_{{\sss T}j}}
\newcommand\etaj{\eta_j}
\newcommand\epg{\epsilon_\gamma}
\newcommand\epc{\epsilon_c}
\newcommand\epem{e^+e^-}
\newcommand{\bfk}{\mbox{\boldmath $k$}}
\newcommand{\pup}{p^\uparrow}
\newcommand{\aup}{a^\uparrow}
\newcommand{\bup}{b^\uparrow}
\newcommand{\cu}{c^\uparrow}
\newcommand{\qup}{q^\uparrow}
\newcommand{\pdown}{p^\downarrow}
\newcommand{\adown}{a^\downarrow}
\newcommand{\bdown}{b^\downarrow}
\newcommand{\cd}{c^\downarrow}
\newcommand{\qdown}{q^\downarrow}
\newcommand\SS{\scriptsize}
\renewcommand\sss{\scriptscriptstyle}
\renewcommand\as{\alpha_{\sss S}}       
\newcommand{\ttbs}{\char'134}
\newcommand{\AmS}{{\protect\the\textfont2
  A\kern-.1667em\lower.5ex\hbox{M}\kern-.125emS}}
\def\beq{\begin{equation}}
\def\eeq{\end{equation}}
\def\eq{\beq\eeq}
\def\beqn{\begin{eqnarray}}
\def\eeqn{\end{eqnarray}}
\def\abs#1{\left| #1\right|}

\clearpage
\section{PREREQUISITES FOR SPIN PHYSICS AT RHIC}\label{sec:PREREQ}
\subsection{Theoretical Concepts and Tools}
\subsubsection{Partons in High-Energy Scattering: Factorization}
Polarized $pp$ collisions at RHIC
will take place at center-of-mass energies of $\sqrt{s}=$200--500 GeV. 
Except for polarization, we have a typical collider 
physics situation, similar to that at CERN's Sp$\bar{\rm p}$S or
the Tevatron at Fermilab. One therefore expects that parton model concepts,
augmented by the predictive power of perturbative QCD, will play a crucial
role in describing much of the interesting spin physics to be studied at RHIC,
if the reaction under consideration involves a hard probe, for
instance a photon produced at transverse momentum ($p_T$) of a few GeV or more.

\begin{figure}[h] 
  \centerline{\psfig{figure=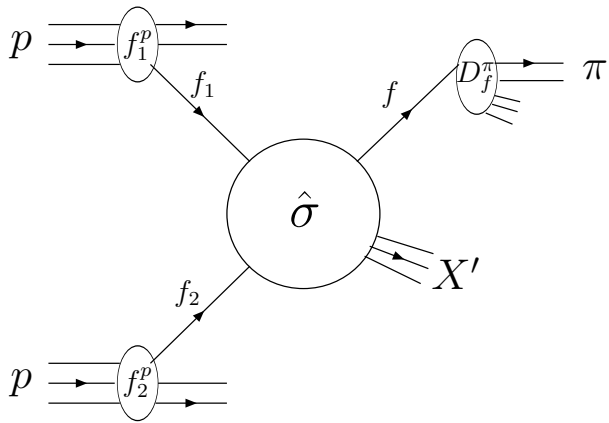,width=8cm,height=5.6cm,angle=0}}
\caption{\label{fig1}\small Production of a large-$p_T$ pion in a 
hard $pp$ collision.}
\end{figure}

The QCD-improved parton model has been successfully applied to many 
high-energy processes involving hadrons in the initial or 
final state. In this framework, a cross section is written in a
factorized form as a convolution of appropriate parton densities and/or 
fragmentation functions with a partonic subprocess cross section. The 
predictive power of perturbative QCD follows from the universality 
of the distribution functions: Once extracted from the data in one 
process, they can be used to make definite predictions for any other. 
As an example, let us consider the production of a pion with large 
$p_T$ in a collision of unpolarized protons, that is, 
$pp\rightarrow \pi X$. The process is depicted in Figure~\ref{fig1}. 
In the parton model framework, in the context of QCD perturbation theory, 
one writes the cross section as a convolution,
\begin{eqnarray} \label{eq1}
\frac{d \sigma^{pp\rightarrow \pi X}}{d{\cal P}} &=& \hspace*{-0.2cm}
\sum_{f_1,f_2,f} \int \hspace*{-0.1cm} 
dx_1 dx_2 dz f_1^p (x_1,\mu^2) f_2^p (x_2,\mu^2)  \nonumber \\ &&
\times \frac{d\hat{\sigma}^{f_1 f_2\rightarrow fX'}}{d{\cal P}} 
(x_1\, p_1,x_2\,p_2,p_{\pi}/z,\mu) D_f^{\pi} (z,\mu^2),  
\end{eqnarray}
where $p_1$ and $p_2$ are the incident proton momenta. Here, ${\cal P}$ stands
for any appropriate set of the kinematic variables of the reaction.
Furthermore, $f_i^p (x,\mu^2)$ is introduced 
as the probability density for finding a parton of type $f_i$ in the proton,
which has taken fraction $x$ of the proton's momentum. Likewise, 
$D_f^{\pi} (z,\mu^2)$ is the probability density for finding a pion with 
momentum fraction $z$ in the parton $f$. The 
$\hat{\sigma}^{f_1 f_2\rightarrow fX'}$
are the underlying hard-scattering cross sections for initial partons
$f_1$ and $f_2$ producing a final-state parton $f$ plus unobserved $X'$. 

The functions $f^p$ and $D_f^{\pi}$ introduced in Equation~\ref{eq1} 
express intrinsic properties of the proton and of the hadronization mechanism, 
respectively. Therefore, they are sensitive to non-per\-tur\-ba\-tive physics 
and cannot be calculated from first principles in QCD at present. In contrast
to this, for a sufficiently hard process, it will make sense to calculate 
the subprocess cross sections $\hat{\sigma}^{f_1 f_2\rightarrow fX'}$ 
as perturbation series in the strong coupling $\alpha_s$.
The separation of short-distance and long-distance phenomena as embodied
in Equation~\ref{eq1} necessarily implies the introduction of an unphysical 
mass scale $\mu$, the factorization scale. The presence of $\mu$ arises 
in practice when computing higher-order corrections to the 
$\hat{\sigma}^{f_1 f_2\rightarrow fX'}$. Here, one encounters singularities 
resulting from configurations in which one of the incoming (massless) partons 
collinearly emits another parton. In the same way, such ``collinear'' 
singularities (or ``mass'' singularities) occur in the final state 
from collinear processes involving parton $f$. Regularization of the mass
singularities always introduces an extra mass scale $M$ to the problem; the
cross section depends on it through powers of ``large'' logarithms of the type 
$\ln (p_T/M)$. The collinear-singular logarithms are separated off at the
factorization scale $\mu$, to be of the order of the hard scale $p_T$ 
characterizing the hard interaction, and are absorbed (``factorized'') into the  
``bare'' parton densities (or fragmentation functions). This procedure is of 
use only if it is universal in the sense that the mass singularities absorbed
into the parton densities are the same for all processes involving a given 
initial parton. Proof of this property is the subject of factorization
theorems~\cite{libby78,collins92a} and is necessary for the parton model 
to be valid in the presence of QCD interactions. 

In summary, the QCD-improved parton-model picture as used for Equation~1  
consists of perturbatively calculable partonic hard-scattering cross
sections and of scale-dependent parton densities and fragmentation 
functions that are universal in the sense that once they are measured in one 
process, they can be used 
to make predictions for any other hard 
process. It is important to point out that the parton densities and 
fragmentation functions are never entirely nonperturbative: Their 
dependence on the factorization scale is calculable perturbatively, once the
densities are known at some initial scale $\mu_0$. This has to be so,
since the $\mu$-dependence of the $\hat{\sigma}^{f_1 f_2\rightarrow fX'}$
is calculable and the prediction of a physical quantity, 
such as the 
hadronic cross section $\sigma^{pp\rightarrow \pi X}$, has to be independent 
of $\mu$ to the order of perturbation theory considered. The tool to 
calculate the dependence of the $f^p$ and $D_f^{\pi}$ on the 
``resolution scale'' $\mu$ is the set of evolution equations~\cite{altarelli77}.
\subsubsection{Spin-Dependent Parton Densities and Cross Sections}
So far we have disregarded the {\em spin} information contained in 
parton distributions and fragmentation functions. If a hard-scattering
process with incoming protons having definite spin orientation is studied,
as at RHIC, one expects it to give information on the spin 
distributions of quarks and gluons 
in a polarized proton. The possible
parton distribution functions~\cite{jaffe91} are summarized in 
Table~\ref{tab1}. A similar table could be presented for polarized 
fragmentation functions~\cite{ji94}: The observation 
of the polarization of a final-state hadron should give information on 
the polarization of the parton fragmenting into that hadron.
\begin{table}[t]
\renewcommand{\arraystretch}{1.6}
\begin{center}
\caption{Compilation of quark and gluon parton densities including 
spin dependence.  
The ubiquitous argument $(x,\mu^2)$ of 
the densities has been suppressed. For brevity, a column for antiquarks 
($\bar{q}$) was omitted, which would have an identical structure to that of the quark column. 
Labels $+,-$ denote helicities, and $\uparrow,\downarrow$ transverse 
polarizations. Superscripts refer to partons and subscripts to the 
parent hadron.}
\vspace*{0.3cm}
\begin{tabular}{|c|c|c|} \hline \hline
Polarization & Quarks & Gluons \\ \hline \hline
unpolarized & $q\equiv q_+^+ + q_+^-\equiv q_{\uparrow}^{\uparrow} + 
q_{\uparrow}^{\downarrow}$ & $g\equiv g_+^+ + g_+^-$ \\ \hline
long. polarized & $\Delta q = q_+^+ - q_+^-$ & $\Delta g = g_+^+ - g_+^-$ 
\\ \hline
transversity & $\delta q = q_{\uparrow}^{\uparrow} - 
q_{\uparrow}^{\downarrow}$ & --- \\ \hline
\hline
\end{tabular}
\vspace*{0.3cm}
\label{tab1}

\vspace*{0.1cm}
\end{center}
\end{table}

Within roughly the past decade, beautiful data~\cite{dsigma} have become 
available that are sensitive to the ``longitudinally'' polarized 
(``helicity-weighted'') parton densities of the nucleon. The tool to 
obtain such information has been deep-inelastic scattering (DIS) of 
longitudinally polarized leptons and nucleons. The 
spin asymmetry measured in such reactions gives information on the 
probability of finding a certain parton type ($f=u,\bar{u},d,\bar{d},\ldots,
g$) with positive helicity in 
a nucleon of positive helicity, minus the probability for finding it 
with negative helicity (see Table~\ref{tab1}). These densities are denoted as 
$\Delta f (x,\mu^2)$. 
The Appendix provides a brief 
discussion of the implications of present polarized DIS data on our 
knowledge about the $\Delta f$. 
Within a parton-model concept, 
the integrals of the $\Delta f (x,\mu^2)$ over all momentum Bjorken-$x$ 
(``first moments''), multiplied by the spin of the parton $f$, will by 
definition give the amount of the proton's spin carried by species $f$,
appearing in the proton-spin sum rule:
\begin{equation} \label{spinsr}
\frac{1}{2} =  \int_0^1 dx  \Bigg[ \frac{1}{2}\sum_q \left(
\Delta q + \Delta \bar{q} \right) (x,\mu^2) \; + \; \Delta g (x,\mu^2) \Bigg]
\; + \; L (\mu^2)\; ,
\end{equation}
where $L$ is the orbital angular momentum of quarks and gluons in the
proton~\cite{sivers89}.

The longitudinally polarized parton distributions in Table~\ref{tab1} 
can be separated from the unpolarized ones if suitable differences
of cross sections for various longitudinal spin settings of the initial
hadrons are taken~\cite{craigie83}:
\begin{eqnarray} \label{eq4}
\hspace*{-2cm}
\frac{d \Delta \sigma^{pp\rightarrow \pi X}}{d{\cal P}} &\equiv&
\frac{1}{4} \Bigg[ \frac{d \sigma_{++}^{pp\rightarrow \pi X}}{d{\cal P}} -
\frac{d \sigma_{+-}^{pp\rightarrow \pi X}}{d{\cal P}} -
\frac{d \sigma_{-+}^{pp\rightarrow \pi X}}{d{\cal P}} +
\frac{d \sigma_{--}^{pp\rightarrow \pi X}}{d{\cal P}} \Bigg]  \nonumber \\
\hspace*{-2cm}
&=& \sum_{f_1,f_2,f} \int dx_1 dx_2 dz \; \Delta f_1^p (x_1,\mu^2) \; 
\Delta f_2^p (x_2,\mu^2) \nonumber \\ 
&&\times \frac{d\Delta \hat{\sigma}^{f_1 f_2 \rightarrow fX'}}{d{\cal P}} 
(x_1, p_1,x_2,p_2,p_{\pi}/z,\mu) \; D_f^{\pi} (z,\mu^2) \; ,  
\end{eqnarray}
where
\begin{equation} \label{eq5}
\frac{d \Delta \hat{\sigma}^{f_1 f_2 \rightarrow fX'}}{d{\cal P}} \equiv
\frac{1}{4} \Bigg[ \frac{d\hat{\sigma}_{++}^{f_1 f_2 \rightarrow fX'}}{d{\cal P}}-
\frac{d\hat{\sigma}_{+-}^{f_1 f_2 \rightarrow fX'}}{d{\cal P}}-
\frac{d\hat{\sigma}_{-+}^{f_1 f_2 \rightarrow fX'}}{d{\cal P}}+
\frac{d\hat{\sigma}_{--}^{f_1 f_2 \rightarrow fX'}}{d{\cal P}} \Bigg] \; .
\end{equation}
Here and in Equation~\ref{eq4} subscripts denote the helicities of the 
incoming particles, i.e.\ of the protons in Equation~\ref{eq4} and of partons
$f_1,f_2$ in Equation~\ref{eq5}. Thus, the ``longitudinally polarized'' cross section
$d \Delta \sigma^{pp\rightarrow \pi X}/d{\cal P}$ depends only\footnote{In 
addition, there is dependence on the pion fragmentation functions 
$D_f^{\pi}$.} on the parton densities for longitudinal polarization and 
on the (calculable) ``longitudinally polarized'' subprocess cross sections 
$d \Delta \hat{\sigma}^{f_1 f_2 \rightarrow fX'}/d{\cal P}$. 
A measurement of $d \Delta \sigma^{pp\rightarrow \pi 
X}/d{\cal P}$ therefore gives access to the $\Delta f$. Adding, on the other 
hand, all terms in the first line of Equation~\ref{eq4}, one simply returns to 
the unpolarized cross section in Equation~\ref{eq1}, with its unpolarized
densities $f$ and the unpolarized subprocess cross sections 
$d \hat{\sigma}^{f_1 f_2 \rightarrow fX'}/d{\cal P}$, corresponding also to
taking the sum of the terms in Equation~\ref{eq5}. 

Notice that we have taken both initial protons to be polarized in 
Equation~\ref{eq4}. If only one is polarized, 
we can still define
a singly polarized cross section by $d \sigma_{-}^{pp\rightarrow \pi X}/
d{\cal P} - d \sigma_{+}^{pp\rightarrow \pi X}/d{\cal P}$, 
where the subscript refers to the polarized proton's helicity. However, this combination 
can be nonzero only if parity is violated in the hard 
process~\cite{craigie83}.
If so, the single-spin cross section will depend on products 
of parton densities $\Delta f_1$ and $f_2$, representing the 
polarized and the unpolarized proton, respectively.

With two transversely polarized beams, one will take the first line 
of Equation~\ref{eq4} for transverse polarizations rather than helicities. 
The result will be a polarized cross section depending on transversely
polarized subprocess cross sections and, for each proton, on the differences 
of distributions of quarks (or antiquarks) with transverse spin 
aligned and anti-aligned with the transverse proton spin. The latter 
quantities are the ``transversity'' 
distributions~\cite{ralston79,jaffe91,ji92,artru90} and are 
denoted
$\delta f(x,\mu^2)$ (see Table~\ref{tab1}).\footnote{One frequently also finds the notation
$\Delta_T f (x,\mu^2)$ or $h_1^f (x,\mu^2)$ in the literature.}  Note that in the case of transverse
polarization a $\cos(2\phi)$ dependence of the cross section on the 
azimuthal angle $\phi$ of the observed final-state particle 
arises~\cite{ralston79,jaffe91,ji92,artru90}, since an extra axis is defined 
by the transverse spin. We also mention that transverse single-spin
cross sections, such as $d \sigma_{\uparrow}^{pp\rightarrow \pi X}/
d{\cal P} - d \sigma_{\downarrow}^{pp\rightarrow \pi X}/d{\cal P}$,
are allowed to be nonzero in QCD but vanish in the simple parton-model
picture presented so far~\cite{pumplin78,sivers90} 
(see Section~\ref{subsec:sspin}).

Extension to polarization in the final state is also possible. 
If the observed particle in Equation~\ref{eq1} were, say, a $\Lambda$-hyperon 
instead of the (spinless) pion, one could consider the first line of 
Equation~\ref{eq4} for the helicities of one of the incoming protons (the 
other proton is assumed to be unpolarized, for simplicity) and of 
the $\Lambda$. In this way one obtains a ``helicity transfer'' cross 
section~\cite{craigie83} that depends on the distribution of parton 
$f_2$ for the unpolarized proton, on $\Delta f_1$ for the polarized 
proton, on polarized fragmentation functions $\Delta D_f^{\Lambda}$
(defined in analogy with $\Delta f$), and on helicity-transfer subprocess 
cross sections. 

\begin{figure}[h] 
\hspace*{-4mm}
\epsfysize10cm
\leavevmode\epsffile{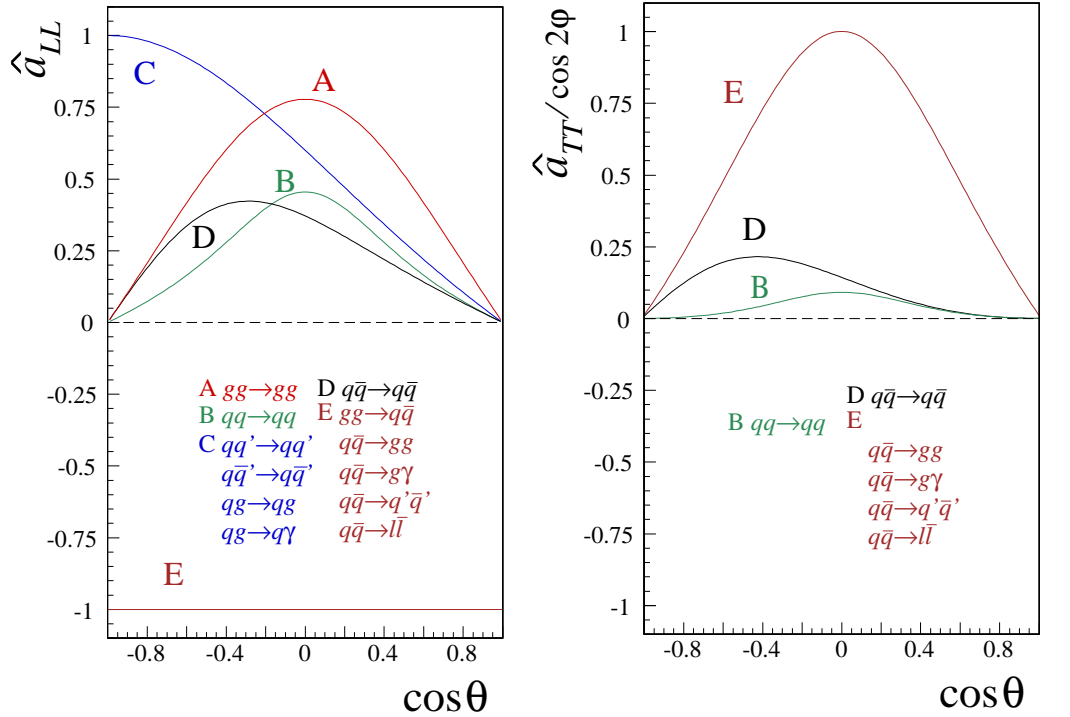}
\caption{\small 
Lowest-order analyzing powers for various reactions
relevant for RHIC, as functions of the partonic center-of-mass system (cms) 
scattering 
angle~\protect\cite{craigie83,jaffe96}. 
%
%
{\it Left:} longitudinal polarization, 
{\it right:} transverse polarization (a factor $\cos(2 \phi)$ has been 
taken out, where $\phi$ is the azimuthal angle of one produced
particle).}
\label{figap}
\end{figure}

For spin experiments, the most 
important quantity in practice is not the polarized cross section itself, 
but the spin asymmetry, which is given by the ratio of the polarized 
over the unpolarized cross section. For our example above, it reads
\begin{equation}
A^{\pi}_{LL}  = \frac{d \Delta \sigma^{pp\rightarrow \pi X}/d{\cal P}}
{d \sigma^{pp\rightarrow \pi X}/d{\cal P}} \; .
\end{equation}
For the asymmetry, one often uses subscripts to denote the type of polarization
($L$=longitudinal, $T$=transverse) of the initial particles.
As follows from Equation~\ref{eq1}, the resulting spin asymmetry will possess 
the generic structure
\begin{equation}
A_{LL} = \frac{ \sum_{f_1,f_2,f} \, \Delta f_1 \times \Delta f_2  \times 
\Big[ d\hat{\sigma}^{f_1 f_2\to fX'}\, \hat{a}_{LL}^{f_1 f_2\to fX'} \Big] 
\times D_f}{ \sum_{f_1,f_2,f} \, f_1 \times f_2  
\times \Big[ d\hat{\sigma}^{f_1 f_2\to fX'} \Big] \times D_f}, 
\end{equation}
where $\hat{a}_{LL}^{f_1 f_2\to fX'}=d\Delta \hat{\sigma}^{f_1 f_2\to fX'}/
d\hat{\sigma}^{f_1 f_2\to fX'}$
is the spin asymmetry for the subprocess $f_1 f_2 \to f X'$, 
often also referred to as 
the analyzing power of the reaction considered. The lowest-order 
analyzing powers for many reactions interesting at RHIC are depicted 
in Figure~\ref{figap}.

\subsection{Detection}
\subsubsection{Asymmetries and Errors}
Asymmetries in a collider experiment can be defined (and measured!) for a single
polarized beam or for both beams polarized, with longitudinally polarized
beams, transversely polarized beams, or with a combination of these.
Additionally, one can study a combination of beam spin state and final-state angular dependence.  For longitudinal polarization for both beams,
the asymmetry $A_{LL}$ is defined 
as
\begin{equation}
A_{LL}=\frac{(\sigma_{++}+\sigma_{--})-(\sigma_{+-}+\sigma_{-+})}
{(\sigma_{++}+\sigma_{--})+(\sigma_{+-}+\sigma_{-+})}.
\label{allsigma}
\end{equation}
Here, $\sigma_{+-}$ represents a cross section for producing a specified final state
with the initial proton helicities ($+$) and ($-$). However, the proton
beams are not in pure helicity states.  We expect that the beams will
be about 70\% polarized, meaning that
\begin{equation}
P_{beam}=\frac{B_+ - B_-}{B_+ + B_-}=0.7,
\label{polarization}
\end{equation}
where $B_+$ refers to the number of protons in the beam with (+) helicity.
Therefore, collisions with two bunches of protons, with for example +0.7
polarization for one bunch and $-$0.7 polarization for the other bunch,
will include collisions of all four helicity combinations, ($++$), ($+-$),
($-+$), and ($--$).   The experimental asymmetry
is defined as follows:
\begin{equation}
A_{LL}=\frac{1}{P_1 P_2}\times \frac{(N'_{++}+N'_{--})-(N'_{+-}+N'_{-+})}
{(N'_{++}+N'_{--})+(N'_{+-}+N'_{-+})},
\label{alln}
\end{equation}
where $N'_{+-}$ represents the observed number of events when the beams
were polarized (+) for beam~1 and ($-$) for beam~2, and normalized by the
luminosity for the crossing. Here, it is 
necessary to know only the {\it relative} luminosity for the ($++$) and ($--$) collisions versus the
($+-$) and ($-+$) collisions.  The beam polarizations are $P_1$ and
$P_2$.  
Algebra can confirm that Equation~\ref{alln}
is equivalent to Equation~\ref{allsigma}.

Similarly, we can define the parity-violating asymmetry for one beam
polarized longitudinally,
\begin{equation}
A_L=-\frac{\sigma_+ - \sigma_-}{\sigma_+ + \sigma_-}~~, \;\;\;
\label{alsigma}
A_L=-\frac{1}{P}\times \frac{N'_+ - N'_-}{N'_+ + N'_-}.
\label{aln}
\end{equation}
The parity-violating asymmetry was defined in 1958 to be positive for left-handed production~\cite{goldhaber58}.  Observed parity-violating
asymmetries are 
therefore typically positive, due to the left-handed weak
interaction.

For transverse spin, one- and two-spin asymmetries are defined in
analogy with the longitudinal asymmetries above, referred to as $A_N$ and
$A_{TT}$.  In this case, the directions ($+$) and ($-$) are transverse
spin directions of the beam protons, not the helicities.  The transverse-spin
asymmetries depend on the production angle, $\theta$, and on the azimuthal
angle of the scattering, $\phi$, as well as other variables.
The azimuthal dependence for scattering two spin-1/2 particles is
\begin{equation}
A_{TT}\propto \cos(2\phi )~~~{\rm and}~~~A_N\propto \cos(\phi).
\label{azdep}
\end{equation}
$\phi$=0 is defined for scattering in the plane perpendicular to the polarization
direction.  Typically the beam is
polarized vertically, with (+) polarization up, and positive $A_N$
implies more scattering to the left than to the right of the beam direction.
The notation $A_{NN}$ is also used for a transverse two-spin asymmetry,
where $N$ refers to beam
polarization normal to the scattering plane. 
A subscript $S$ traditionally designates beam
polarization in the transverse direction in the scattering plane.

 From Equation~\ref{alln} or Equation~\ref{aln} we need to know the beam
polarization(s), count the number of signal events for each combination
of beam spin directions, and monitor the relative luminosity for the
crossings with these combinations of beam spin directions.  The statistical
error of the measurement is
\begin{equation}
(\Delta A_{LL})^2=\frac{1}{NP_1^2P_2^2}-\frac{1}{N}A_{LL}^2.
\label{staterror-all}
\end{equation}
Here $N$ is the total number of events observed, and it is assumed that the
statistical errors on the relative luminosities and on the beam polarization
are small.  For the single spin asymmetry,
\begin{equation}
(\Delta A_{L})^2=\frac{1}{NP_1^2}-\frac{1}{N}A_{L}^2.
\label{staterror-al}
\end{equation}
For small to moderate asymmetries,
\begin{equation}
\Delta A_{LL}=\pm 1/(P_1 P_2)\times
\frac{1}{\sqrt{N}}~~{\rm and}~~\Delta A_L=\pm 1/P\times \frac{1}{\sqrt{N}}.
\label{staterror}
\end{equation}
Since we expect $P=P_1=P_2=0.7$, $\rm{10^4}$ events would give
an error of $\Delta A_{LL}=\pm 0.02$ for the double spin asymmetry, or
$\Delta A_L=\pm 0.014$ for the parity-violating asymmetry.

\begin{figure}[h]
\begin{center}
\hspace*{-0.9cm}
\epsfysize3cm
\leavevmode\epsffile{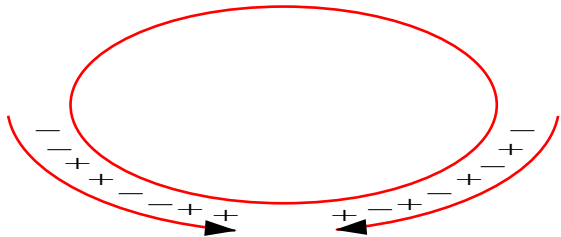}
\caption{\label{F:Bunch}
\small Bunch filling pattern with
respect to the spin states of polarized protons.}
\end{center}
\end{figure}

In principle, asymmetry measurements are very straightforward.
As long as the detector acceptance remains stable with time between
reversals of the beam spin states, the measurement will be stable and the
errors will be largely statistical.  However, when reversals of the beam
polarization are spread apart in time, and/or the beam  conditions for
opposite spin states differ, acceptance can change and false asymmetries
develop.
At RHIC the  bunches, 120 in each ring, are prepared independently at the
source, so that the bunches can alternate polarization sign, 106~ns apart, as
shown in
Figure~\ref{F:Bunch}.  Note that one ring with alternate bunches and the other
ring with alternating pairs of bunches create the four spin combinations,
($++$), ($+-$), ($-+$), and ($--$).  Therefore, the concern of time-dependent
acceptance and  beam location variations for opposite sign beams should
be negligible at RHIC,  and asymmetry measurement errors should be mainly
statistical, even for small asymmetries.

What systematic errors do we expect at RHIC?  There are two classes of
systematic errors:  false asymmetries and scale errors.  If the relative
luminosities for the bunch spin combinations are incorrectly measured
through, for example, a saturation effect in the luminosity monitor, which
couples to variations in beam intensity for the bunch spin combinations,
the numerator of Equation~\ref{alln} or Equation~\ref{aln} will be nonzero from
the incorrect normalization, creating a false asymmetry.  If the beam
polarization is incorrect, no false asymmetry is created, but the scale
of the resulting asymmetry is changed.

Each experiment will measure the relative luminosities for each crossing.
The luminosity monitors must be independent of beam
polarization, and statistical errors on the relative luminosity
measurements need to be very small to match the statistical
sensitivity available for high-statistics measurements, such as jet production.

Relative
luminosity needs to be known to the ${\rm 10^{-4}}$ level for some
asymmetry measurements.  This job appears daunting, but the time dependence
of the acceptance (efficiency is included with acceptance in this discussion)
of the luminosity monitor needs to be stable only over roughly one turn of RHIC,
or 13~$\mu$s.

\subsubsection{Polarimetry}
Polarization is measured by using a scattering process with known
analyzing power.  Knowledge of the analyzing power for different processes
can come from theoretical calculations, for example for QED processes,
and from experimental measurements using a beam or target with known
polarization.  Polarimetry at RHIC~\cite{alekseev} will be based on
elastic proton-proton
and elastic proton-carbon scattering in the Coulomb nuclear interference
(CNI) region.  The analyzing power there is largely calculable; it is expected
to be small but significant. It can be determined to excellent precision using a polarized proton
target in RHIC, and the rates for CNI scattering are
very high.

Sensitivity to the proton spin is from scattering the Coulomb field of an
unpolarized particle (proton or carbon) from the magnetic
moment of the polarized proton.  This method uses the dominance of the
interference of the one-photon exchange helicity-flip electromagnetic
amplitude,
proportional to the proton anomalous magnetic moment, with the
non-flip strong hadronic amplitude, which is determined by the $pp$ or
$pC$ total cross
section $\sigma_{\rm tot}$~\cite{schwinger,BGL,kz,trueman,bs}.  However,
there can also
be a hadronic spin-flip term, which is not presently calculable.
(This possibility is discussed in more detail in Section~7.)
Therefore, significant sensitivity to the proton spin is predicted over the
entire RHIC energy range from the electromagnetic term, but the absolute
sensitivity is limited to $\pm$15\%~\cite{trueman}.  For this reason,
a polarized hydrogen gas jet target will be installed at RHIC.  The
polarization of the jet
target can be measured to $\pm3\%$ so that the analyzing power
in the CNI region can be measured precisely, and this analyzing power
will then be used to determine the beam polarization at RHIC precisely.

Existing polarized hydrogen gas jet targets are thin, so that the determination
of the beam polarization using the jet target will take hours.  For this
reason, RHIC will also use carbon ribbon targets and use proton-carbon
CNI scattering to monitor the beam polarization frequently.

Absolute beam polarization is expected to be known to $\pm 5\%$.
The systematic scale uncertainty of the asymmetry measurements will be of the order of
$\pm 5\%$ for single spin measurements such as Equation~\ref{aln} and
$\pm 10\%$ for two spin measurements such as Equation~\ref{alln}.
By scale uncertainty we mean that in forming the ratio of the error in
the asymmetry measurement, $\Delta A_{LL}$ in Equation~\ref{staterror}, to
the measurement itself, $A_{LL}$ in Equation~\ref{alln}, the polarization
normalization divides out.  Therefore, the polarization uncertainty
applies to the scale of the measurement and not to the statistical
significance of the measurement.

\subsubsection{RHIC Detectors}
This article emphasizes the physics that will be probed at RHIC-Spin.
There are six collision points at RHIC, as shown in Figure~\ref{F:RHICSPIN},
and two are used for the two large detectors,
{\sc Phenix}~\cite{Morrison97} and {\sc Star}~\cite{Harris94}. These
detectors are quite complementary: {\sc Star}
emphasizes large coverage with tracking, and the strengths of {\sc Phenix}
are in fine-grained calorimetry for photons and electrons
and in ``forward" muon detectors.  Sensitivities for the spin measurements at RHIC are
based on these detectors.  Although one could discuss the sensitivity for
a $4\pi$-acceptance fine-grained detector, such a detector does not exist.
And we note that, for example, the
{\sc Phenix} electromagnetic calorimeter (EMCal) has 100 times finer
granularity than previous collider detectors.  The {\sc pp2pp}~\cite{Guryn}
and {\sc Brahms}~\cite{Videbaek94} detectors share one collision point,
and the {\sc Phobos}~\cite{Baker95}
detector is located at another crossing. {\sc Star} and {\sc Phenix} will
measure gluon and quark polarizations with hard scattering. The {\sc pp2pp}
experiment will measure spin dependence in small-angle elastic scattering;
{\sc Brahms} and {\sc Phobos} will measure transverse spin asymmetries.

\begin{figure}[h]
  \centerline{\psfig{figure=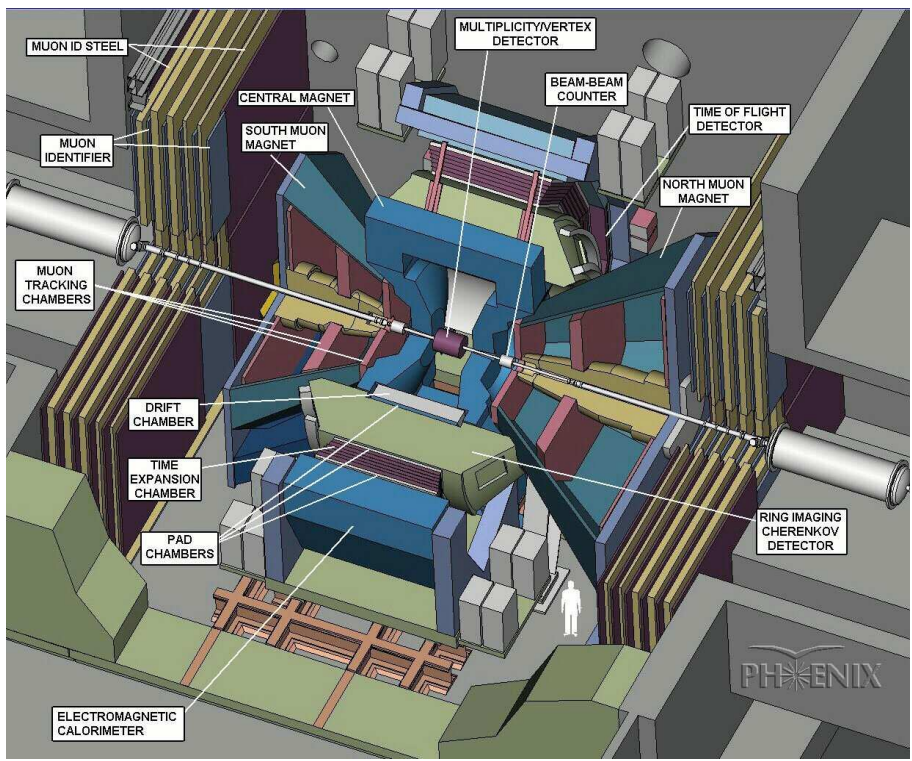,width=12cm,height=10cm,angle=0}}
\caption{\label{F:PHENIX}
\small The {\sc Phenix} detector system.}
\end{figure}

The {\sc Phenix} detector,
shown in Figure~\ref{F:PHENIX}, has two central
arms at $\rm{90^{\circ}}$ to the beams with fine-grained EMCal
towers, $\Delta\eta\times\Delta\phi = 0.01\times~0.01$.
The minimum opening angle for $\pi^0\rightarrow\gamma\gamma$ corresponds
to one tower for a 30-GeV $\pi^0$.  This is important to separate directly
produced photons, a probe of gluon polarization, from background from $\pi^0$
decay.   Resolution is excellent, with $\Delta E/E=\pm 3\%$
at 10~GeV.  The two central arms each cover $\rm{90^{\circ}}$ in azimuth,
left and right.  Pseudorapidity acceptance is $|\eta |<0.35$.
The vertex detector, central tracker, ring-imaging Cherenkov detector
(RICH), and time expansion chamber (TEC) are also shown. The central
magnetic field, provided by two Helmholtz coils, is 0.8 Tesla meters,
integrated radially. The tracking 
$p_T$ resolution is
$\Delta p_T/p_T=\pm 2.5\%$ at 10~GeV/$c$ for the east arm, which includes
the TEC, and $\pm 5\%$ in the west arm without a TEC.  Triggering in
the central arms, for selection of high-$p_T$ direct photons, electrons,
and charged pions, will be based on overlapping tower clusters in the
EMCal, combined with RICH information.  Studies indicate a sufficiently
clean and efficient electron trigger to allow $p_T>1~$GeV/$c$ or so.
Such a low-momentum electron trigger is attractive to obtain charm
quark events.

\begin{figure}[h]
  \centerline{\psfig{figure=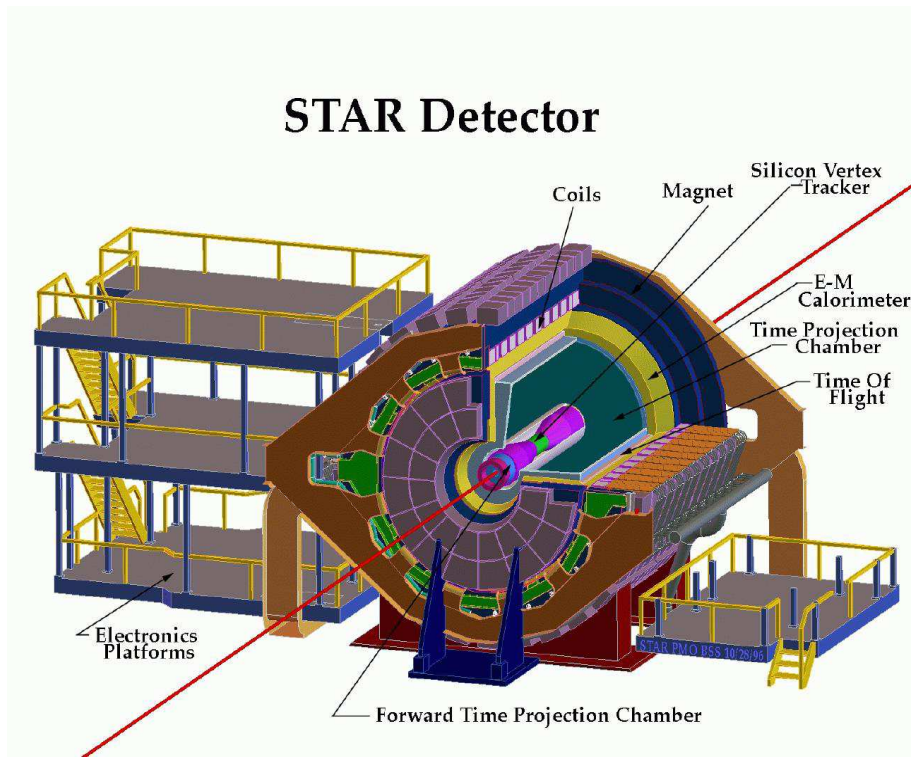,width=12cm,height=10cm,angle=0}}
\caption{\label{F:STAR}
\small Cut view of {\sc Star} detector system.}
\end{figure}

The {\sc Phenix} muon arms surround the beams, covering $\Delta\phi = 2\pi$
and $1.2<|\eta|<2.4$. The arms include a muon identifier (MuID)
with five iron-detector layers, as well as
three tracking stations.  The muon arm magnets produce a radial field,
ranging from 0.2 to 0.75 Tesla meters, integrated along the beam direction.
Longitudinal momentum resolution is about $\pm 2\%$ at 10~GeV/$c$.
A 4-GeV/$c$ muon  penetrates to the
fifth MuID layer.

{\sc Phenix} will emphasize muon measurements for $W\rightarrow \mu\nu$,
Drell-Yan lepton pairs, $J/\psi$, and heavy flavors.  Central arms
will measure
$\gamma$, jet fragmentation to $\pi^{0,\pm}$, and $W\rightarrow e\nu$,
as well as heavy flavors (single lepton, and $e$ with $\mu$), with small acceptance
and high granularity.

The {\sc Star} detector is shown schematically in Figure~\ref{F:STAR}.
A barrel time projection chamber (TPC) covers $|\eta|<1.0$ and $\Delta
\phi=2\pi$.  This is surrounded by an EMCal with towers
$\Delta\eta\times\Delta \phi = 0.05\times 0.05$.
The EMCal has a shower maximum detector at 
a depth of five radiation lengths with
projective readout wire chambers reading longitudinally and azimuthally, each
with 1-cm spacing. Studies show an effective separation of
single photons from merged photons from
$\pi^0$ decay out to $p_T$ = 25~GeV/$c$. Energy resolution is
excellent, with $\Delta E/E=\pm$ 5\% at 10~GeV.  Additional barrel
detection includes a silicon drift vertex
tracker around the collision point and an array of
trigger counters outside the TPC.
The central solenoid
field is 1.0 Tesla meters, integrated radially.
The {\sc Star} 
$p_T$ resolution is
$\pm$3\%  at $p_T$ = 10~GeV/$c$.
{\sc Star} is also building one endcap calorimeter to cover
$1<\eta<2$ for photons and electrons and to expand the jet cone coverage.

Triggering at {\sc Star} will be based on the trigger counters and
EMCal, which are fast detectors.  A major issue
to resolve is the long memory of the TPC, which will include on the
order of 800 out-of-time
tracks from the 40-$\mu$s drift time, at full luminosity with
a 10-MHz collision
rate.  {\sc Star} studies have shown that the drift of the out-of-time
tracks cause them to
point significantly away from the collision point.  This drift
will be used to remove
the unwanted tracks, and this must be done before writing to
tape.  Studies have also
shown good jet reconstruction after the tracks are
removed.  Jets will be reconstructed
at {\sc Star} with a combination of EMCal and tracking, with no hadronic
calorimetry.  Simulations show a
full width at half maximum of 30\% for the $\Delta p_{jet}/p_{jet}$
distribution, limited by the hadronization dynamics of final-state partons.
A cone size of $R=\sqrt{\Delta\eta^2+\Delta \phi^2}=0.7$ was
used. {\sc Star} will measure jets,
$\gamma + jet$, and $W\rightarrow e\nu$, with wide acceptance and
reconstruction of the parton kinematics.

The {\sc pp2pp} experiment is designed to study small-angle proton-proton
elastic scattering, from $-t=0.0005$ to $-t=1.5~({\rm GeV}/c)^2$.
Silicon-strip detectors will be placed in Roman pots
at two locations along each beam to measure scattering to very small angles.
The experiment 
will also use a polarized hydrogen jet target with
silicon recoil detectors to cover lower center-of-mass energy and 
will measure the absolute polarization of the RHIC beams.

The {\sc Brahms} detector has two movable spectrometers
($2.3^{\circ} \le \theta  \le 30^{\circ}$ and
$30^{\circ} \le \theta  \le 95^{\circ}$) with superb particle
identification.
The spectrometer covers up to 30~GeV/$c$ with
$\Delta p/p=\pm 0.1$\% and it will provide unique
measurements of single transverse-spin asymmetries in the forward,
thus high-$x_F$, region.

{\sc Phobos} is a table-top--sized detector that uses 
silicon-strip detectors to cover a large solid angle.
Two spectrometers comprise
15 planes of silicon pad detectors each, with 7 planes in a 2-Tesla
magnetic field. Its wide geometrical acceptance
and momentum resolution is suitable for pair
or multiparticle final states in spin physics, such as
$\rho^{0} \rightarrow \pi^{+} \pi^{-}$.

\clearpage
\section{MEASURING $\Delta g$ AT RHIC}\label{sec:dg}
\begin{figure}[htb]
\centerline{
   \psfig{figure=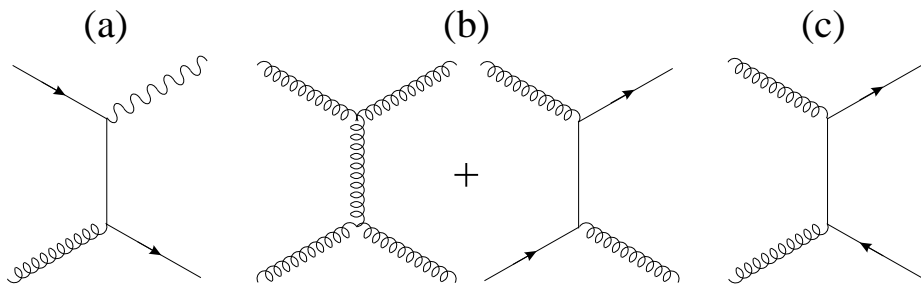,width=0.9\textwidth,clip=}
           }
\caption{ 
Selected lowest-order Feynman diagrams for elementary processes 
with gluons in the initial state in $pp$ collisions:
($a$) quark-gluon Compton process for prompt-photon production, 
($b$) gluon-gluon and gluon-quark scattering for jet production, 
and ($c$) gluon-gluon fusion producing a heavy quark pair.}
\label{F:Gluon}
\end{figure}

Measurement of the gluon polarization in a polarized proton is a major
emphasis and strength of RHIC-Spin. By virtue of the spin sum 
rule~(\ref{spinsr}), a large $\Delta g$ is an exciting possible 
implication~\cite{bass99} of the measured~\cite{dsigma}
smallness of the quark and antiquark contribution to the proton spin. A large 
gluon polarization would imply unexpected dynamics in the proton's spin 
structure. Because of this special importance of $\Delta g$, and since
it is left virtually unconstrained by the inclusive-DIS experiments performed 
so far (see Appendix), several experiments focus on its measurement.
A fixed-target DIS experiment, {\sc Hermes}, measures the process 
$\vec{e}(\vec{\gamma}) \vec{p} \rightarrow h^+ h^- X$~\cite{hermes99}, 
where $h=\pi,K$, which is in principle sensitive to the gluon polarization.  
However, the transverse momenta are low, making interpretation in a 
hard-scattering formalism difficult.  The DIS 
experiment {\sc Compass}~(see e.g.\ \cite{mallot96}) will measure the same reaction
at higher energies, as well as heavy-flavor production, to access 
gluon polarization. Scaling violations and the reaction $\vec{e} 
(\vec{\gamma})  \vec{p} \rightarrow {\rm jet(s)} X$ will constrain 
$\Delta g$ at HERA, if the proton ring is  polarized~\cite{deroeck97}. 
At RHIC, the gluon polarization will be measured directly, 
precisely,
and over a large range of gluon momentum fraction, with large momentum
transfer ensuring the applicability of perturbative QCD to describe the
scattering, and with several independent processes.
The RHIC probes, shown in Figure~\ref{F:Gluon},
are as follows: 
\begin{itemize} 
\item High-$p_T$ (``prompt'') photon production 
$\vec{p}\vec{p}\rightarrow \gamma X$
\item Jet production, $\vec{p}\vec{p}\rightarrow {\rm jet(s)} X$
\item Heavy-flavor production, $\vec{p}\vec{p}\rightarrow 
c\bar{c}X,b\bar{b}X$
\end{itemize}

\subsection{Prompt-Photon Production}
Prompt-photon production, $pp,p\bar{p},pN\rightarrow 
\gamma X$~\cite{whalley97}, has been {\em the} classical tool for 
determining the unpolarized gluon density at intermediate and large 
$x$. At leading order, a photon in the final state is produced in the 
reactions $qg\to\gamma q$ (Figure~\ref{F:Gluon}$a$) and 
$q\bar{q}\to\gamma g$. Proton-proton, as opposed to proton-antiproton, 
scattering favors the quark-gluon Compton process, since the
proton's antiquark densities are much smaller than the quark ones.
The analyzing power for direct photon production is large (Figure~\ref{figap}).
Photons produced in this way through partonic hard scattering
show a distinct signal at colliders, that of an isolated 
single photon without jet debris nearby. The production of photons 
with polarized beams at RHIC is therefore a very promising method
to measure $\Delta g$~\cite{berger89,indu91,cheng91,bourrely91}. 

If parton kinematics can be approximately reconstructed, one can bin 
the events in the parton momentum fractions $x_1,x_2$ of the hard 
scattering. Assuming dominance of the Compton process, the asymmetry 
$A_{LL}$ for prompt-photon production can then be written as 
\begin{equation} \label{eqqg}
A_{LL} \approx \frac{\Delta g(x_1)}{g(x_1)} \, \cdot \, 
\Bigg[\frac{\sum_q e_q^2 \left[ \Delta q (x_2) + 
\Delta \bar{q} (x_2) \right]}{\sum_q e_q^2 \left[q (x_2)+
\bar{q} (x_2) \right] } \Bigg] \, \cdot \, 
\hat{a}_{LL}(gq \rightarrow \gamma q) + (1\leftrightarrow 2) \, .
\end{equation} 
As a result of the quark charge-squared weighting, 
the second factor in Equation~\ref{eqqg} coincides, to lowest 
order,
with the spin asymmetry $A_1^p$ measured in polarized DIS, 
and the partonic asymmetry $\hat{a}_{LL}$ is calculable in perturbative QCD. 
Thus, from the measurement of $A_{LL}$, one can directly extract 
$\Delta g(x)/g(x)$.

Both {\sc Phenix} and {\sc Star} intend to use this procedure for a
direct leading-order determination of $\Delta g$, where one exploits the
dominance of $2\rightarrow 2$ ($ab\to \gamma c$) 
parton scattering when reconstructing
$x_1,x_2$. This is done either on average based on the detector 
acceptance for the photon, or event-by-event by observing photon-plus-jet 
events ({\sc Star}). Estimates of the ``background'' from 
$q\bar{q}$ annihilation have been made~\cite{bland99}. 
Eventually, the aim will be 
%
a ``global'' QCD analysis 
of polarized prompt photon, and 
other RHIC and DIS, asymmetry data to determine the full set of 
polarized parton densities simultaneously, as is done routinely in 
the unpolarized case~\cite{martin98,cteq5,grv}. In this case, one can 
directly work from the spin asymmetries, and inclusion of, for instance, 
higher-order corrections is more readily possible. 
  
\begin{figure}[htb]
\centerline{
   \psfig{figure=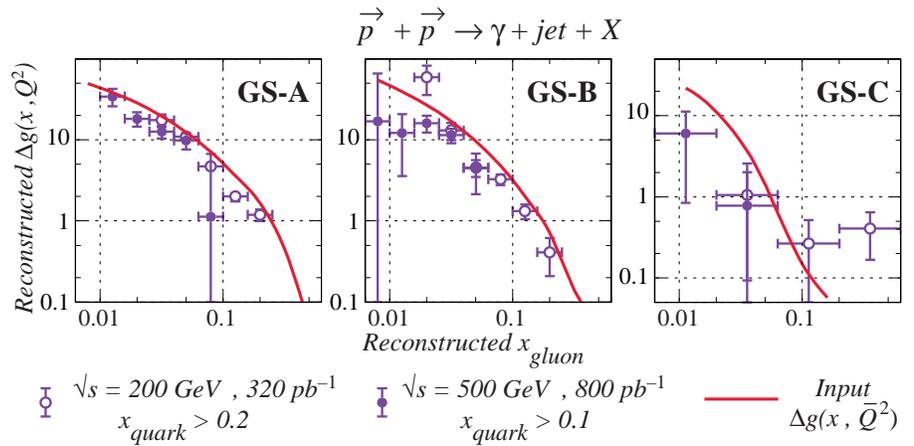,width=0.9\textwidth,clip=}
           }
\caption{ 
Sensitivity of {\sc Star} measurements of $\Delta g(x)$ in the channel
$\vec{p}\vec{p} \rightarrow \gamma + {\rm jet} + X$. }
\label{F:STAR-DG}
\end{figure}

Figure~\ref{F:STAR-DG} shows the level of accuracy {\sc Star} can 
achieve~\cite{bland99} in a direct measurement of $\Delta g$ based 
on reconstructing parton kinematics in photon-plus-jet events. 
The solid lines show 
in each plot the input density employed for $\Delta g(x)$, taken 
from~\cite{gehrmann96}. The data points and the error bars show 
the reconstructed $\Delta g(x)$ and its precision for 
standard luminosities in 
runs at $\sqrt{s}=200$~GeV (open circles) and $\sqrt{s}=500$~GeV 
(solid circles).

High-$p_T$ photons can also be produced through
a fragmentation process, in which a parton, scattered or produced in
a QCD reaction, fragments into a photon plus a number of hadrons.
The need for introducing a fragmentation contribution is physically 
motivated 
by the fact that a QCD hard-scattering process may produce, 
again through a fragmentation process, a $\rho$ meson that has the same 
quantum numbers as the photon and can thus convert into a photon, 
leading to the same signal. In addition, at higher orders, the 
perturbatively calculated direct component contains divergencies 
from configurations where a final-state quark becomes collinear to the 
photon. These singularities naturally 
introduce the need for nonperturbative fragmentation functions into which
they can be absorbed. So far, the photon fragmentation 
functions are insufficiently known; 
information is emerging 
from the LEP experiments~\cite{buskulic96a}. Note that all QCD 
partonic reactions contribute to the fragmentation component;
thus, the benefit of having a~priori only one partonic reaction 
($q\bar{q}\to\gamma g$) competing with the signal ($qg\to\gamma q$) 
is lost, even though some of the subprocesses relevant to the 
fragmentation part at the same time result from a gluon initial state.
Theoretical studies~\cite{aurenche93,gluck94,vogt95,aurenche99} 
for photon production in unpolarized collisions, based on 
predictions~\cite{aurenche93,gluck93,bourhis98} for the photon-fragmentation functions that are compatible with the sparse 
LEP data, indicate that the fragmentation component is in practice 
a small, albeit nonnegligible, effect. 

In the fixed-target regime, fragmentation photons are 
believed~\cite{vogt95} to contribute at most $20\%$ to the direct 
photon cross section. At collider energies, the fragmentation 
mechanism is estimated to produce about half of the observed 
photons; however, 
an ``isolation'' cut can be imposed on the photon 
signal in experiment.  Isolation is an experimental necessity: In a 
hadronic environment, the study of photons in the final state is complicated 
by the abundance of $\pi^0$s, which decay into pairs of $\gamma$s. 
If two photons are unambiguously detected in an event, 
their invariant mass can indicate whether they resulted from a $\pi^0$ 
(or $\eta$) decay. However, 
either escape of one of the decay photons from the 
detector or 
merging of the two photons from 
$\pi^0$ decay at high $p_T$ fake a single photon event. 
The isolation cut reduces this background, 
since $\pi^0$s are embedded in jets. If a given neighborhood of the 
photon is free of 
energetic hadron tracks, it is less likely that the observed photon came
from $\pi^0$ decay, and the event is kept; it is rejected otherwise. 
Traditionally, isolation is realized by drawing a cone of fixed aperture 
in $\varphi$--$\eta$ space around the photon [where $\varphi$ is 
the photon's azimuthal angle and $\eta=-\ln \tan (\theta/2)$ is its 
pseudorapidity, defined through its polar angle $\theta$], and by 
restricting the hadronic transverse energy allowed in this cone to a 
certain small fraction of the photon transverse energy. In this way, 
the fragmentation contribution to single $\gamma$s, resulting from 
an essentially collinear process, will also be 
diminished~\cite{berger91}. It is not expected~\cite{gluck94,vogt95} 
that fragmentation will remain responsible for more than 15--20\% of 
the photon signal after isolation.  
%
It has been suggested~\cite{frixione98} that allowing proportionally less 
hadronic energy, the closer to the photon it is deposited, rather than permitting 
a fixed fraction in the full isolation cone, would improve isolation 
by reducing the fragmentation photons still further .

Several early theoretical studies for isolated prompt-photon production at 
polarized RHIC have been published
(e.g.\ \cite{berger89,indu91,cheng91,bourrely91}). 
The QCD corrections to the direct (i.e.\ 
nonfragmentation) component of polarized prompt-photon production 
were first calculated in References~\cite{contogouris93,gordon93} and are now 
routinely included in theoretical studies 
(e.g.\ \cite{contogouris97,gordon94,gordon97,gordon98,frixione99}). 
In particular, References~\cite{gordon97,gordon98,frixione99} present 
Monte Carlo codes for the next-to-leading-order (NLO) 
corrections to the direct part of the 
cross section,
which 
allow the isolation 
constraints to be taken into account and also have the flexibility 
to predict photon-plus-jet observables, $\vec{p}\vec{p} \to \gamma+jet+X$.
We also emphasize that much effort has gone, and is still going, into 
event-generator studies~\cite{bland99,goto99,gullenstern95,om99}
for prompt-photon physics at RHIC. 

\begin{figure}[h]
\centerline{
   \psfig{figure=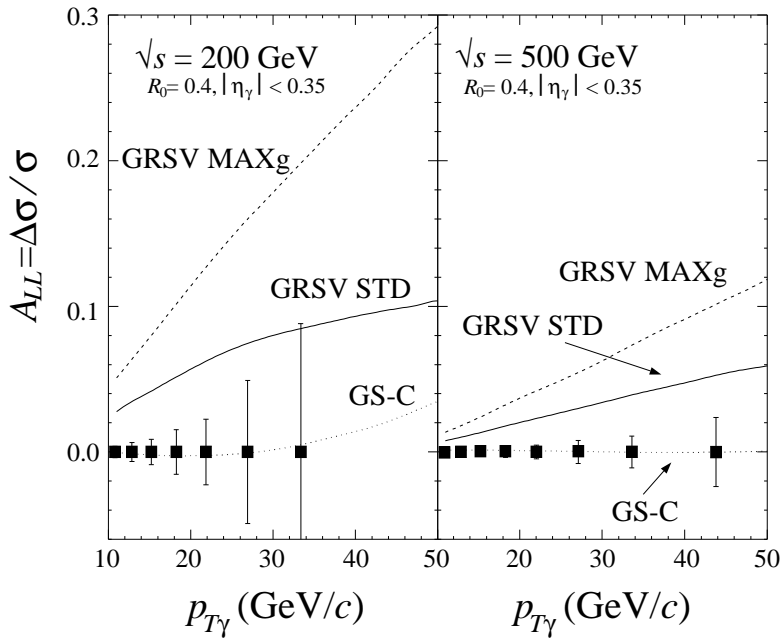,angle=0,width=0.8\textwidth,clip=}
           }
\caption{  
Asymmetry as a function of transverse momentum, for various polarized parton densities,
at different cms energies~\protect\cite{frixione99}. The expected statistical errors for the 
{\sc Phenix} experiment are also shown. 
%
%
}
\label{fig:asypt}
\end{figure}                                                              

Figure~\ref{fig:asypt} shows the asymmetry as obtained in an NLO theory 
calculation, as a function of the photon's transverse momentum $p_T$. 
A rapidity cut $\abs{\eta}<0.35$ has been applied, matching the
acceptance of the {\sc Phenix} experiment. In the left (right) part of the 
figure we plot the asymmetries obtained at $\sqrt{s}=200$~GeV (500 GeV).
The isolation of Reference~\cite{frixione98} was used, with isolation cone 
opening $R_0=0.4$ and $\epg=1,n=1$ (see Reference~\cite{frixione98} 
for details on the latter parameters). The solid, dashed, and dotted 
lines correspond to the NLO predictions obtained with GRSV-STD, 
GRSV-MAXg~\cite{gluck96}, and GS-C~\cite{gehrmann96} polarized parton
densities, respectively. These densities are all compatible with
present data from polarized DIS and differ mainly in their gluon
content: GRSV-MAXg has a very sizeable positive gluon distribution, 
whereas GS-C has a small, and oscillating, $\Delta g$. The gluon of GRSV-STD
lies between the other two. The three gluon densities are shown
in Figure~\ref{figPDF} in the Appendix. The error bars represent the 
expected statistical accuracy for the measurement at {\sc Phenix}, 
with $\Delta\phi =\pi$ and for standard luminosities and beam polarizations.

It is a striking feature of  Figure~\ref{fig:asypt} that different 
spin-dependent gluon densities do indeed lead to very different 
spin asymmetries for prompt-photon production. RHIC experiments
will be able to measure $\Delta g$.  

For fixed $p_T$, higher-energy
probes lower $x$ in the parton distributions, and this leads 
to the smaller predicted asymmetries for $\sqrt{s}$=500~GeV.
If one considers the same $x_T=2p_T/\sqrt{s}$ value for the two 
energies in Figure~\ref{fig:asypt}, the parton densities are 
being probed at similar momentum fractions but rather different 
``resolution'' scales, of the order of $p_T$. It will be interesting 
to see whether measurements performed at different cms 
energies will yield information that is consistent, and compatible,
with QCD evolution.

Present comparisons between theory and experiment
[and possibly between experiment and experiment~\cite{aurenche99}] regarding 
unpolarized direct $\gamma$ production are unsatisfactory~\cite{abe94}. 
Transverse smearing of the momenta of the initial partons participating 
in the hard scattering, substantially larger than that already introduced 
by the NLO calculation, has been considered~\cite{huston95,martin98,kimber99} 
to reconcile theory with data. This approach is partly based on measured 
values of dimuon, dijet, and diphoton pair transverse momenta $k_T$ in 
hadronic reactions~\cite{huston95} and has enjoyed some phenomenological 
success. More recently, the role of all-order-resummations of large
logarithms in the partonic cross section, generated by (multiple) soft-gluon 
emission, has been investigated in the context of prompt-photon 
production~\cite{lai98,laenen98,catani99,laenen00}. Threshold 
resummations~\cite{laenen98} have been shown~\cite{catani99} to lead to 
improvements in the fixed-target regime, and a very recent new 
formalism~\cite{laenen00} that jointly incorporates threshold and $k_T$
resummations has the potential of creating the substantial enhancements
needed for bringing theory into agreement with data. It is likely that 
a better understanding of the prompt-photon process will have been 
achieved by the time RHIC performs the first measurements of 
\mbox{$\vec{p}\vec{p}\rightarrow \gamma X$}. Also, the main problems reside
in the fixed-target region; at colliders there is much less reason 
for concern.  RHIC itself should also be able to provide new and 
complementary information in the unpolarized case---never before have 
prompt-photon data been taken 
in $pp$ collisions at energies as high as \mbox{$\sqrt{s}=$200--500~GeV}.

Finally, we note that it was also proposed~\cite{indu91,sridhar93,berger99} to 
determine $\Delta g$ through the reaction $qg\rightarrow \gamma^{*}q$,
which is again the Compton process, but now with a photon off-shell
by the order of a few GeV and giving rise to a Drell-Yan lepton 
pair of comparable 
$p_T$. The advantage is a cleaner
theoretical description; for instance, no photon fragmentation
component is present in this case. However, compared to 
prompt-photon production at a given $p_T$, the event rate is
reduced by 2--3 orders of magnitude due to the additional factor 
$\alpha_{em}/(3\pi Q^2)$ in the Drell-Yan cross section, where $Q$ 
is the dilepton mass. Higher statistics are available at lower $p_T$,
but at the price of reduced asymmetry and higher background from 
$q\bar{q}\rightarrow \gamma ^* g$ annihilation.

\subsection{Jet Production}
Toward the higher end of RHIC energies, jets could be the key
to $\Delta g$: at $\sqrt{s} =500$~GeV, clearly structured jets will be
copiously produced, and jet observables will show a strong 
sensitivity to $\Delta g$ thanks to the 
dominance~\cite{bourrely91,soffer98} of the 
$gg$ and $qg$ initiated subprocesses (see Figure~\ref{F:Gluon}$b$) in accessible 
kinematical ranges. Jet studies will be performed by {\sc Star}. 
One can alternatively look for high-$p_T$ leading hadrons such as
$\pi^0,\pi^{\pm}$, whose production proceeds through the same partonic 
subprocesses but involves an explicit fragmentation function in the 
theoretical description. This is planned for the {\sc Phenix} experiment,
where the limitation in angular coverage precludes jet studies.

Knowledge of the NLO QCD corrections is expected to be particularly 
important for the case of jet production, since it is only at NLO 
that the QCD structure of the jet starts to play a role in the 
theoretical description, providing for the first time the possibility 
of realistically matching the procedures used in experiment in order to group final-state
particles into jets.
The task of calculating the NLO QCD corrections to polarized jet 
production has been accomplished~\cite{deflorian99}. Furthermore,
a Monte Carlo code that had been designed by Frixione~\cite{frixione97}, based
on Reference~\cite{frixione96} and the subtraction method
in hadron-hadron unpolarized collisions, was extended 
to the polarized case in Reference~\cite{deflorian99}. We emphasize that 
in the unpolarized case, the comparison of NLO theory predictions with 
jet production data from the Tevatron is extremely 
successful~(see e.g.\ \cite{bab00}).

\begin{figure}[htb]
\begin{center} 
\vspace{0.5cm}
\hspace*{-6mm}
\psfig{figure=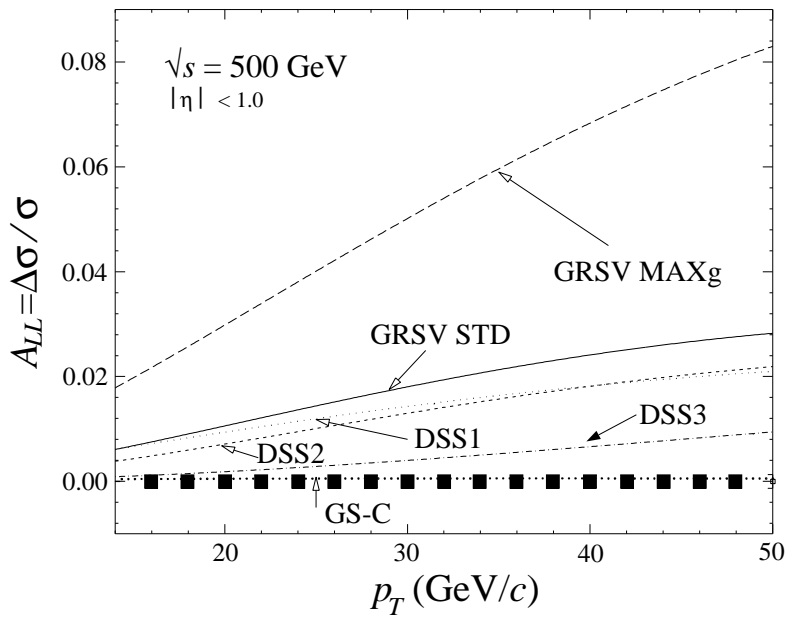,angle=0,width=0.8\textwidth,clip=} 
\end{center} 
\caption{Asymmetry versus jet transverse momentum~\protect\cite{deflorian99} for various 
polarized parton density sets. The "data point" for $p_T$=48~GeV/$c$ 
indicates the
statistical accuracy expected for the {\sc Star}
experiment for standard polarization and luminosity.  Expected errors for lower
$p_T$ are smaller than the points shown.}
\label{jetpt}
\end{figure}

Figure~\ref{jetpt} shows the double-spin asymmetry 
for single-inclusive jet production at NLO as a function of the jet 
$p_T$ and for various polarized parton 
densities~\cite{gehrmann96,gluck96,deflorian98} with different $\Delta g$
(see Figure~\ref{figPDF} in the Appendix). A cut $\abs{\eta}<1$ 
has been applied, and we have chosen the Ellis-Soper (ES) cluster
jet algorithm~\cite{ellis93} with the resolution parameter $D=1$. The 
renormalization and factorization scales have been chosen as $\mu_0\approx 
p_T$ (for further details, see Reference~\cite{deflorian99}). 
The asymmetry shows a strong sensitivity to $\Delta g$. However, the asymmetry 
is rather small, regardless of the specific parton densities used. 
Fortunately, the expected statistical accuracy of such a jet measurement, 
calculated for standard luminosity and indicated in the figure, is very good.

\begin{figure}[htb]
\centerline{
   \psfig{figure=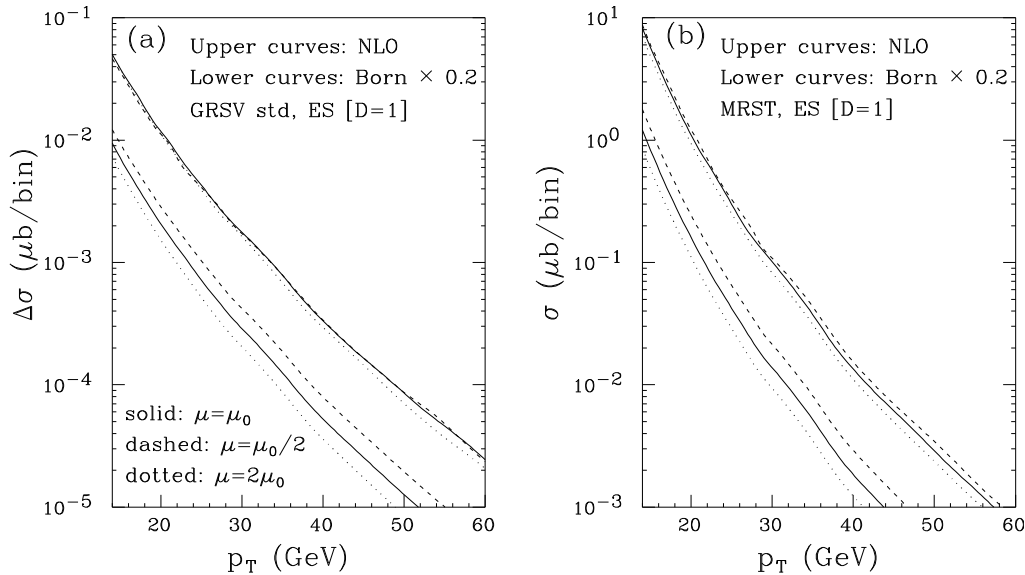,width=1.\textwidth,clip=} }
\caption{ 
   Scale dependence of the 
next-to-leading-order and Born 
   $p_{\sss T}$ distributions for jet production~\protect\cite{deflorian99}.
   ($a$) Polarized $pp$ scattering
   and ($b$) unpolarized $pp$ scattering at 
   $\protect\sqrt{s} =$ 500~GeV. The range
   of the pseudorapidity is  restricted to $|\eta| < 1 $.}
\label{jetscale}
\end{figure}

The inclusion of NLO corrections in jet production, as shown 
in Figure~\ref{jetscale}, leads to a clear reduction in scale 
dependence of the cross section. One thereby
gains confidence that 
it is possible to calculate reliably
the cross section and the spin asymmetry for a given $\Delta g$.
This reduction in scale
dependence after NLO corrections is also seen for direct photon 
production~\cite{frixione99}.

\subsection{Heavy-Flavor Production} 
The production of heavy quark pairs in hadronic collisions
is dominated by gluon-gluon fusion, $gg \rightarrow Q \bar{Q}$
(see Figure~\ref{F:Gluon}$c$). For $pp$ collisions, the competing 
channel $q\bar{q} \rightarrow Q \bar{Q}$ is particularly suppressed, 
since it requires an antiquark in the initial state. Thus, heavy 
quarks provide direct access to the gluons in the proton. 
Early predictions~\cite{contogouris90} 
at the 
lowest order demonstrated that indeed 
this reaction could be used to measure $\Delta g$ in polarized $pp$ 
collisions. The importance of NLO corrections for a quantitative 
analysis was pointed out~\cite{karliner93}. Presently, only the NLO QCD 
corrections to heavy-flavor production in polarized 
photon-photon~\cite{kamal97} and
photon-proton~\cite{bojak98,merebashvili99} collisions are known; 
it is anticipated that the full set of NLO corrections relevant
for polarized $pp$ collisions will be available soon~\cite{Bojak99}. 
It should be mentioned that in the unpolarized case, theoretical NLO 
predictions for hadro- and photoproduction of heavy flavors 
often 
fail to provide a satisfactory description of the 
data~(see \cite{mangano97} for review).

Heavy-flavor production can be selected by 
the channels $pp \rightarrow \mu^{\pm} X$, $pp \rightarrow e^{\pm} X$, 
$pp \rightarrow \mu^+\mu^- X$, $pp \rightarrow e^+ e^-  X$, and
$pp \rightarrow \mu^{\pm} e^{\mp} X$.  
Like sign leptons are also possible from bottom,
with one direct $b$-decay to a lepton and one sequential decay 
through charm. 
Charm and bottom
events will probe the gluon density at different momentum fractions 
and scales, and also enter the analysis with different,
albeit calculable, weights. Experimentally it may be possible to
determine the fraction of the charm production 
rate by, for example, looking at the
channel $pp \rightarrow \mu^+ D^0 X$.

The production of heavy quarkonia is another potentially attractive probe 
of the gluon density with a clear experimental signature.  However, so 
far we do not understand the production mechanism. Predictions 
for $\psi$ production based on the color-singlet
model~\cite{berger81} fall short of experimental data
taken at the Tevatron~(see e.g.\ \cite{abe92}). This has stimulated the development
of a more general approach that also gives rise to potentially 
important color-octet contributions~\cite{bodwin95}. Theoretical studies for
the spin asymmetry in charmonium production in $pp$ collisions have 
been presented~\cite{hidaka80,contogouris90,morii95,teryaev97,jaffe99}.
Reference~\cite{morii95} considers the color-singlet mechanism; 
Reference~\cite{teryaev97} also examines color-octet contributions. 
Sensitivity to the production mechanism as well as to $\Delta g$
is found. Similarly, $\chi_2 (3556)$ production at RHIC would have the 
potential to discriminate between the color-singlet and 
the color-octet mechanisms, as well as 
to pin down $\Delta g$~\cite{jaffe99}. Here, one would have to look at the angular
distribution of the decay photon in $\chi_2 \rightarrow J/\psi + 
\gamma$. The number of observed events for this reaction will unfortunately 
be low at RHIC.

\clearpage
\section{QUARK AND ANTIQUARK HELICITY DISTRIBUTIONS}
\label{wasy}
Measurements in polarized DIS~\cite{dsigma}, when combined with information 
from baryon octet $\beta$-decays~\cite{bass99}, show that the total
quark-plus-antiquark contribution to the proton's spin, summed over all
flavors, is surprisingly small. In the standard interpretation of
the $\beta$-decays~\cite{bass99}, this finding is equivalent to
evidence for a large negative polarization of strange quarks in the 
proton, which makes it likely that also the $SU$(2) ($u,d$) sea 
is strongly negatively polarized. This view is corroborated by the fact 
that in this analysis the spin carried, for example, by $u$ quarks 
comes out much smaller than generally expected in quark models~\cite{bass99}, 
implying that a sizeable negative $u$-sea polarization partly compensates 
that of the valence $u$ quarks. 
Alternative treatments of the information 
from $\beta$-decays~\cite{lipkin,gluck96}, when combined with the DIS results,
also directly yield large negative $\bar{u}$ and $\bar{d}$ polarizations. 
Inclusive DIS (through $\gamma^*$ exchange) itself is 
sensitive 
to the combined contributions of quarks and antiquarks of each flavor 
but cannot provide information on the polarized quark and antiquark 
densities separately (see Appendix). Directly measuring the individual 
polarized antiquark distributions is therefore an exciting task and
will also help to clarify the overall picture 
concerning DIS and the $\beta$-decays. 

Further motivation for dedicated measurements of antiquark densities 
comes from unpolarized 
physics. Experiments in 
recent years have shown~\cite{NMC91,baldit94,hermesu}
a strong breaking of $SU$(2) symmetry in the antiquark sea, 
with the ratio $\bar{d}(x)/\bar{u}(x)$ rising to 1.6 or higher. It is very
attractive to learn whether the polarization of $\bar{u}$ and $\bar{d}$
is large and asymmetric as well. RHIC experiments will measure the 
$\bar{d}/\bar{u}$ unpolarized ratio and the $\bar{u}$ and $\bar{d}$ 
polarizations separately.

Semi-inclusive DIS measurements~\cite{adeva96} are one approach to 
achieving a separation of quark and antiquark densities. This method
combines information from proton and neutron (or deuteron) targets
and uses correlations in the fragmentation process between the type 
of leading hadron and the flavor of its parton progenitor, expressed 
by fragmentation functions. The dependence on the details of the
fragmentation process limits the accuracy of this method. 
At RHIC the polarization of the $u,\bar{u},
d,$ and $\bar{d}$ quarks in the proton will be measured directly and 
precisely using maximal parity violation for production of $W$
bosons in $u\bar{d}\rightarrow W^+$ and $d\bar{u}\rightarrow W^-$ \cite{craigie83,BOURRELY93,bourrely95,kamal98,gehrmann98}.
In addition, at RHIC, inclusive production of 
$\pi$, $K$, and $\Lambda$ will be used to measure quark and
antiquark polarization through the fragmentation process. 
Another probe at RHIC will be Drell-Yan 
production of lepton 
pairs~\cite{was85,cheng91,bourrely91,mathews92,ratcliffe,kamal98,gehrmann98}.

\subsection{Weak Boson Production} 
Within the standard model, $W$ bosons are produced through pure $V$-$A$ interaction. 
Thus, the helicity of the participating quark and antiquark are fixed 
in the reaction. In addition, the $W$ couples to a weak charge that
correlates directly to flavors, if we concentrate on one generation. 
Indeed the production of $W$s in $pp$ collisions is dominated by 
$u,d, \bar{u}$, and $\bar{d}$, with some contamination from 
$s, c, \bar{s}$, and $\bar{c}$, mostly through quark mixing. Therefore 
$W$ production is an ideal tool to study the spin-flavor structure 
of the nucleon. 

\begin{figure}[h]
  \centerline{\psfig{figure=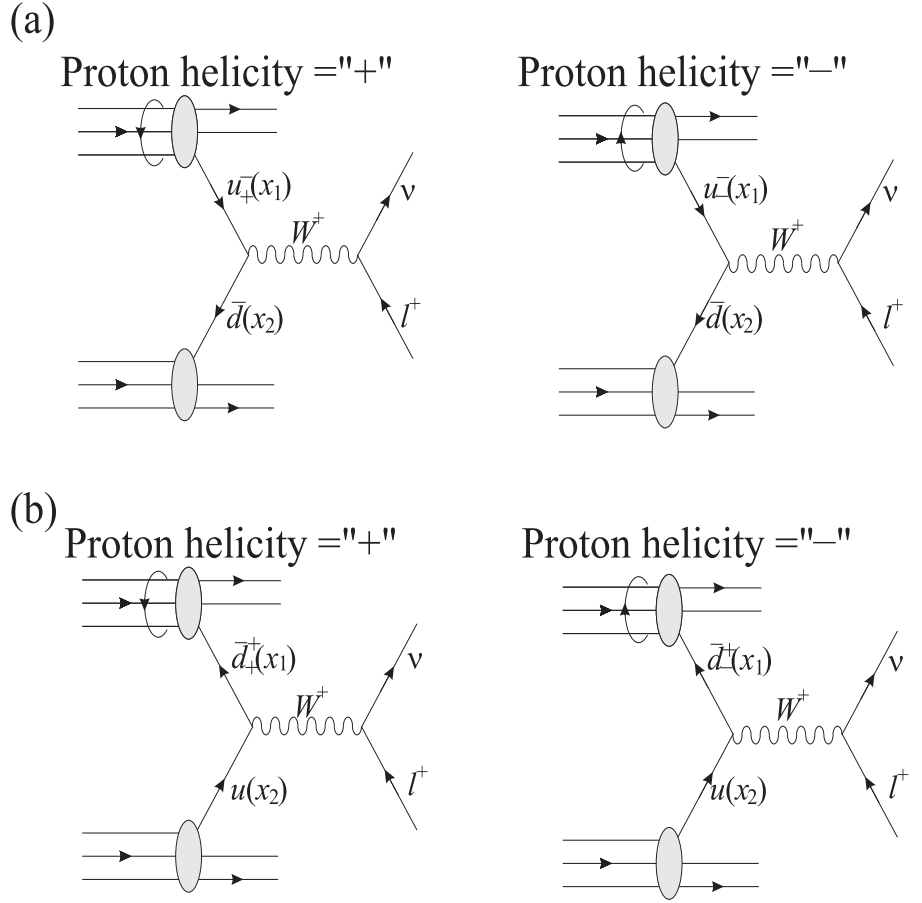,width=12cm,height=12cm,angle=0}}
\caption{\small Production of a $W^+$ in a $\vec{p}p$ collision, at lowest
order. ($a$) $\Delta u$ is probed in the polarized proton. ($b$) $\Delta 
\bar{d}$ is probed.}
\label{F:AL_W}
\end{figure} 
The leading-order production of $W$s, $u\bar{d}\rightarrow W^+$, 
is illustrated in Figure~\ref{F:AL_W}.  The longitudinally
polarized proton at the top of each diagram collides with an unpolarized
proton, producing a $W^+$.  
At RHIC the polarized protons will be
in bunches, alternately right- ($+$) and left- ($-$) handed.  
The parity-violating
asymmetry is the difference of left-handed and right-handed production of
$W$s, divided by the sum and normalized by the beam polarization:
\begin{equation}
A_L^W = \frac{1}{P} \times \frac{N_-(W)-N_+(W)}{N_-(W)+N_+(W)}~~.
\label{alw}
\end{equation}
As Figure~\ref{F:Bunch} shows, we can construct this asymmetry
from either polarized beam, and by summing over the helicity states of the 
other beam. The production of the left-handed weak bosons violates 
parity maximally. Therefore, if for example the production of the $W^+$ 
proceeded only through the diagram in Figure~\ref{F:AL_W}$a$, the 
parity-violating asymmetry would directly equal the 
longitudinal polarization asymmetry of the $u$ quark in the 
proton:
\begin{equation}
A_L^{W^+} = \frac{u^{-}_{-}(x_1) \bar{d}(x_2) - u^{-}_{+}(x_1) \bar{d}(x_2)}
                 {u^{-}_{-}(x_1) \bar{d}(x_2) + u^{-}_{+}(x_1) \bar{d}(x_2)}
          =  \frac{\Delta u(x_1)}{u(x_1)}.
\label{alw+u}
\end{equation}
Similarly, for Figure~\ref{F:AL_W}$b$ alone,
\begin{equation}
A_L^{W^+} = \frac{\bar{d}^{+}_{-}(x_1) u(x_2) - \bar{d}^{+}_{+}(x_1) u(x_2)}
                 {\bar{d}^{+}_{-}(x_1) u(x_2) - \bar{d}^{+}_{+}(x_1) u(x_2)}
          =  -\frac{\Delta\bar{d}(x_1)}{\bar{d}(x_1)}.
\label{alw+bard}
\end{equation}
In general, the asymmetry is a superposition of the two 
cases:
\begin{equation}
A_{L}^{W^+} 
= \frac{\Delta u(x_1) \bar{d}(x_2) - \Delta \bar{d} (x_1) u (x_2) } 
{u(x_1) \bar{d}(x_2) + \bar{d} (x_1) u (x_2) }.
\end{equation} 
To obtain the asymmetry for $W^-$, one interchanges $u$ and $d$. 

For the $pp$ collisions at RHIC with $\sqrt{s}=500$~GeV, the quark will
be predominantly a valence quark. By identifying the rapidity of the $W$, 
$y_W$, relative to the {\it polarized} proton, we can 
obtain direct measures of the quark and antiquark 
polarizations, separated by quark flavor: $A_L^{W^+}$ approaches 
$\Delta u/u$ in the limit of $y_W \gg 0$, whereas for 
$y_W \ll 0$ the asymmetry becomes  $-\Delta \bar{d}/\bar{d}$.
Higher-order corrections change the asymmetries only a
little~\cite{kamal98,gehrmann98}. 

The kinematics of $W$ production and Drell-Yan production of 
lepton pairs is the same. The momentum fraction carried by the 
quarks and antiquarks, $x_1$ and $x_2$ (without yet assigning
which is which), can be determined from $y_W$,
\begin{equation}
x_1=\frac{M_W}{\sqrt{s}}e^{y_W},~~~x_2=\frac{M_W}{\sqrt{s}}e^{-y_W}.
\label{E:x-W}
\end{equation} 
Note that this picture is valid for the predominant 
production of $W$s at $p_T=0$. The experimental difficulty is that the 
$W$ is observed through its leptonic decay $W \rightarrow l \nu$, 
and only the charged lepton is observed. 
We therefore need to relate the lepton kinematics to $y_W$, so that we can 
assign the probability that the polarized proton provided the quark 
or antiquark. Only then will we be able to translate the measured 
parity-violating asymmetry into a determination of the quark or antiquark 
polarization in the proton. 

The rapidity of the $W$ is related to the lepton rapidity in the $W$ rest 
frame ($y_l^{*}$) and in the lab frame ($y_{l}^{{\rm lab}}$) by
\begin{equation} 
y_{l}^{lab} = y_{l}^{*} + y_W,~~{\rm where}~~
y_{l}^{*}= \frac{1}{2}{\rm ln}\Bigg[ \frac{1+{\rm cos}\theta ^{*} }
{1-{\rm cos}\theta ^{*} } \Bigg] \; .
\end{equation} 
Here $\theta^{*}$ is the decay angle of the lepton in the $W$ rest 
frame, and cos$\theta^{*}$ 
can be determined from the transverse momentum ($p_T$) of the lepton with an 
irreducible uncertainty of the sign~\cite{bland00}, since
\begin{equation} 
p_{T}^{\rm lepton} =p_T^{*} = \frac{M_W}{2} {\rm sin}\theta^{*}. 
\end{equation}  
In this reconstruction, the 
$p_T$ of the $W$ is neglected. 
In reality, it has a $p_T$, resulting for example from higher-order 
contributions such as $gu \rightarrow W^{+}d$ and $u\bar{d} 
\rightarrow W^{+}g$, or from primordial 
$p_T$ of the
initial partons. 

Usually $W$ production is identified by requiring charged leptons 
with large 
$p_T$ and large missing transverse 
energy, due to the undetected neutrino. Since none of the detectors at RHIC 
is hermetic, measurement of missing 
$p_T$ is not 
available, 
which leads to some background. 
Possible sources of leptons with high 
$p_T$
include charm, bottom, and vector boson production. Above 
$p_T\ge 20$~GeV/$c$, leptons from $W$ decay dominate, with 
a smaller contribution from $Z^0$ production. 
Both {\sc Phenix} and {\sc Star} can estimate the single-lepton 
$Z^0$ background from measured $Z^0$ production. 
The additional background from misidentified hadrons is 
expected to be small.

\begin{figure}[h]
  \centerline{\psfig{figure=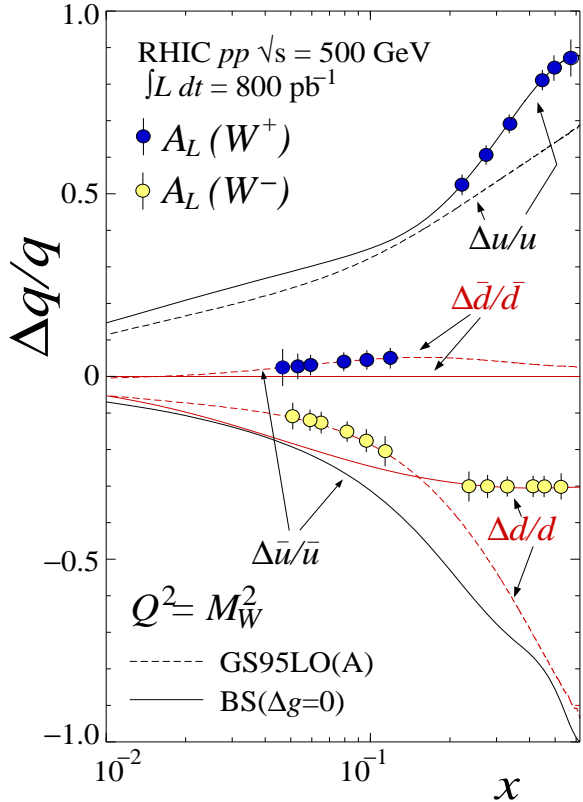,width=8cm,height=11cm,angle=0}}
\caption{\small Expected sensitivity for the flavor-decomposed
quark and antiquark polarization overlayed on the parton densities
of Reference~\protect\cite{bourrely95} (BS) and of 
Reference~\protect\cite{gehrmann96} [GS95LO(A)].
Darker points and error bars refer to the sensitivity 
from $A_{L}(W^+)$ measurements, and lighter ones correspond to $A_{L}(W^-)$.}
\label{F:W_polpdf}
\end{figure}

Expected yields were estimated with {\sc Pythia}~\cite{SJOSTRAND94} 
and {\sc ResBos}~\cite{BALAZS97}. The cross section at RHIC for $W^+$ 
($W^-$) production is about 1.3~nb (0.4~nb). These estimates vary 
by 5--10\% according to the choice of the parton 
distribution set. For 800~pb$^{-1}$ and $p_T\ge 20$~GeV/$c$, {\sc Phenix} 
expects about 8000 $W^+$s and 8000 
$W^-$s in the muon arms (that the numbers 
are equal is due to the decay angle distribution and acceptance), 
as well as 15,000 $W^+$ and 2500 $W^-$ electron decays in the central 
arms. {\sc Star}, with its large acceptance for electrons, expects 72,000 
$W^+$s and 21,000 $W^-$s. Using Equation~\ref{E:x-W} to reconstruct $x$, 
Figure~\ref{F:W_polpdf} shows the expected sensitivity for $\Delta f(x)/f(x)$, 
with $f=u,d,\bar{u},\bar{d}$, for the {\sc Phenix} muon data.

\begin{figure}[h] 
  \centerline{\psfig{figure=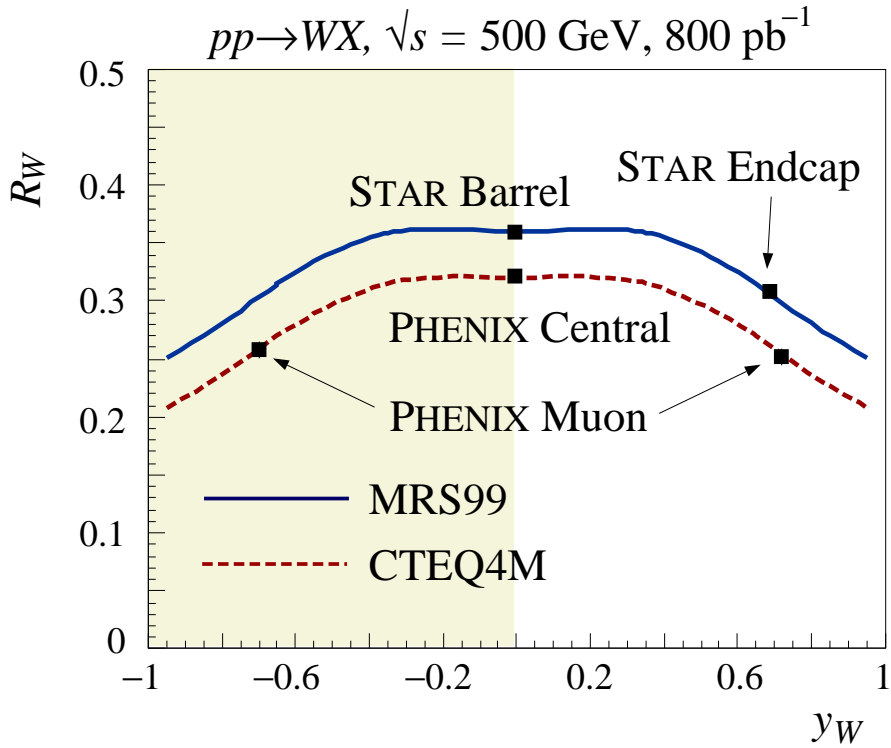,width=12cm,height=10cm,angle=0}}
\caption{\label{F:R_W}{\small
The ratio $R_W=(d\sigma(W^-)/dy)/(d\sigma(W^+)/dy)$ for unpolarized 
$pp$ collisions at RHIC. The shaded region indicates that unpolarized
$pp$ collisions are symmetric in $y_W$.  
To illustrate the sensitivity of the 
measurement, we show an earlier set of 
parton densities (CTEQ4M~\protect\cite{cteq4}) and a 
set (MRS99~\protect\cite{martin99}) that 
includes the latest information from Drell-Yan 
data~\protect\cite{baldit94}.
Both curves include an asymmetric sea with $\bar{d}/\bar{u}$
rising to 1.6 for increasing antiquark 
momentum fraction $x_{\bar{q}}$, but the latter also
includes a drop-off in the ratio for higher $x_{\bar{q}}$.}}
\end{figure}

RHIC will also significantly contribute to our knowledge about the 
unpolarized parton densities of the proton, since it will have the 
highest-energy $pp$ collisions. $\bar{p}p$ production of $W$s has
a much stronger valence component in the determined~\cite{abewasy}
$u(x)/d(x)$ ratio. Isospin dependence in Drell-Yan production of 
muon pairs in $pp,pd$ scattering~\cite{baldit94}, 
violation of the Gottfried sum rule~\cite{GOTTFRIED67,NMC91}, and recent 
semi-inclusive DIS measurements~\cite{hermesu} have shown that the 
unpolarized sea is not $SU$(2) symmetric.
At RHIC, the ratio of unpolarized $W^+$ and $W^-$ cross sections
will directly probe the $\bar{d}/\bar{u}$ ratio, as shown in 
Figure~\ref{F:R_W}.

\subsection{Drell-Yan Production of Lepton Pairs}
Drell-Yan production of lepton pairs has been a basis for information about 
sea quarks~\cite{baldit94}. At lowest order, lepton pairs are created from 
quark-antiquark annihilation. With knowledge of the quark densities, Drell-Yan
cross sections give the antiquark distributions versus $x$. The spin 
asymmetry $A_{LL}$ for Drell-Yan lepton pair production in collisions 
of longitudinally polarized proton beams is proportional to 
a sum of contributions over quark flavors, each a product of the 
polarized quark density times the antiquark distribution. The subprocess 
analyzing power is maximally negative, $\hat{a}_{LL}=-1$. 
One therefore has, at lowest order,
\begin{equation}
A_{LL} = \hat{a}_{LL}\times
 \frac{\sum_q e_{q}^{2} \{ \Delta q(x_1) \Delta \bar{q}(x_2)
                     + \Delta \bar{q}(x_1) \Delta q(x_2) \} }  
                     {\sum_q e_{q}^{2} \{ q(x_1) \bar{q}(x_2)
                     + \bar{q}(x_1) q(x_2) \} }. 
\end{equation}
This asymmetry is parity-conserving if the process proceeds via a 
photon. Since the cross sections by flavor are weighted by the 
electric charge squared, the asymmetry is dominated by the 
$u\bar{u}$ combination and gives information on the $\bar{u}$ 
polarization, with the $u$ quark polarization as input.
NLO corrections to the asymmetry have been calculated~\cite{ratcliffe,kamal98}
to be small for low  $p_T$ of the virtual photon. 
For higher $p_T$, Drell-Yan production is sensitive to $\Delta g(x)$ 
through $qg \rightarrow \gamma^{*} q$~\cite{indu91,sridhar93,berger99}, as 
discussed in Section~\ref{sec:dg}. 

However, lepton pair production in high-energy $pp$ collisions 
is dominated by coincidental semileptonic decays of heavy-quark pairs, 
e.g.\ $b \rightarrow c \, l^- \, \bar{\nu} $ in the low-mass
region. The feasibility of the measurements will therefore depend on the 
ability to separate or estimate this background. Estimates of the yields 
in the {\sc Phenix} muon arms obtained with {\sc Pythia} for $pp$ 
collisions at $\sqrt{s}=$200~GeV show that lepton pairs with invariant 
mass $M \ge$6~GeV/$c^2$ are dominated by Drell-Yan production. One 
expects $\sim$40,000 pairs for a nominal integrated luminosity 
of 320~pb$^{-1}$. 

\clearpage
\section{TRANSVERSE AND FINAL-STATE SPIN EFFECTS}\label{sec:TRANS}
Exciting physics prospects also arise for transverse polarization
of the RHIC proton beams. 
One is the possibility of a first
measurement of the quark transversity densities 
introduced
in Table~\ref{tab1}. The transversity distributions, measuring 
differences of probabilities for finding quarks with transverse 
spin aligned and anti-aligned with the transverse nucleon spin, 
are 
as fundamental as the longitudinally polarized densities
for quarks and gluons, $\Delta q$, $\Delta g$;
they have evaded measurement so far 
because
they decouple from inclusive DIS. Comparisons of the polarized quark
distributions $\delta q$
and $\Delta q$ are particularly interesting; in the nonrelativistic
limit, where boosts and rotations commute, one has 
$\delta q(x,Q^2)=\Delta q(x,Q^2)$. Deviations from this provide
a measure of the relativistic nature of quarks inside
the nucleon.

Studies of single-transverse spin asymmetries, defined similarly 
to Equation~\ref{aln}, will be a further interesting application. They
arise as ``higher-twist'' effects (that is, they are suppressed by 
inverse powers of the hard scale) and probe quark-gluon correlations 
in the nucleon. They have an exciting history in experiments that 
were carried out at energies much lower than RHIC's, where
large polarizations and single-spin asymmetries have been seen~\cite{bunce76}. 
Yet another field of 
spin physics to be thoroughly examined by the RHIC experiments 
will be the transfer of longitudinal or transverse polarization 
from the initial into the final state, which then leaves traces 
in the polarization of hadrons produced in the fragmentation process.
\subsection{The Quark Transversity Distributions}
\begin{figure}[b]
\hspace*{1cm}
\psfig{figure=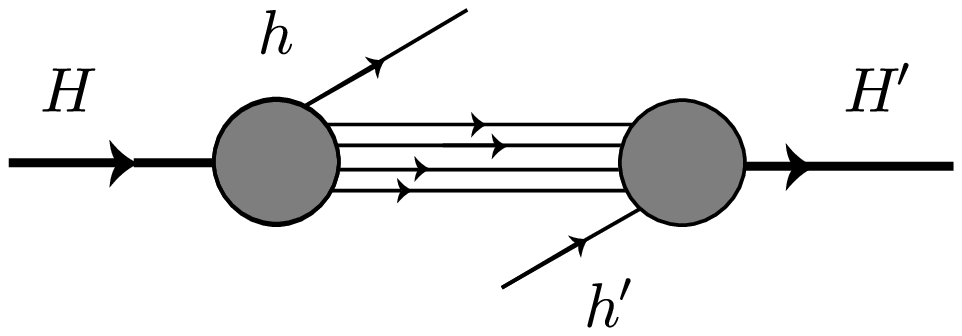,width=5.5cm}
\hspace*{1cm}
\psfig{figure=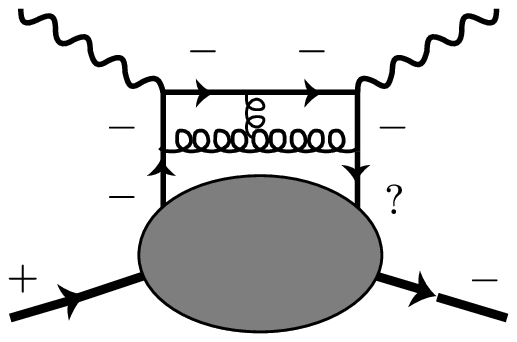,width=4cm}
\caption[]{\small {\it Left:} quark densities as related to
polarized quark-hadron forward scattering in the $u$-channel. Labels
refer to helicities. {\it Right:} decoupling of transversity from 
deep-inelastic scattering~\cite{jaffe97a}. Quark
chirality is not changed by coupling to a photon or a gluon.}
\label{ffig0} 
\end{figure}

The transversity densities $\delta q$ and $\delta \bar{q}$ are virtually 
inaccessible in inclusive DIS~\cite{jaffe91,artru90}. We can see this as
follows~\cite{goldstein95}. In a simple parton model, and working in a
helicity basis, we can view the quark densities as imaginary parts of
%
%
polarized quark-hadron forward scattering in the $u$-channel,
denoted by ${\cal A} (H,h;H',h')$ (see Figure~\ref{ffig0}). 
One then has $q= {\cal A}(++;++) + {\cal A}(+-;+-)$, 
$\Delta q= {\cal A}(++;++) - {\cal A}(+-;+-)$, but
$\delta q= {\cal A}(++;--)$. Thus, for transversity to contribute, 
the quark has to undergo a helicity flip in the hard scattering,
which is not allowed (for massless quarks) at the DIS quark-photon
vertex due to helicity conservation. Note the striking feature that the 
helicity labels of the final state in ${\cal A}(++;--)$ differ from 
those of the initial state. In other words, the complex conjugate amplitude 
contained in ${\cal A}(++;--)$ refers to a different physical state than 
the 
initial. This ``off-diagonal'' nature in terms of helicity 
is usually referred to as chiral-odd~\cite{jaffe91} and
can indeed in practice only be achieved by having transverse 
polarization, which can be written as a superposition of helicity states.

Another important consequence is that, unlike the situation for 
unpolarized and longitudinally polarized densities, there is no 
transversity gluon distribution~\cite{jaffe91,ji92,artru90}. This is due 
to angular momentum conservation; a gluonic helicity-flip amplitude 
would require the hadron to absorb two units of helicity, which a
spin-$1/2$ target cannot do.
 
\begin{figure}[h]
\centerline{\psfig{figure=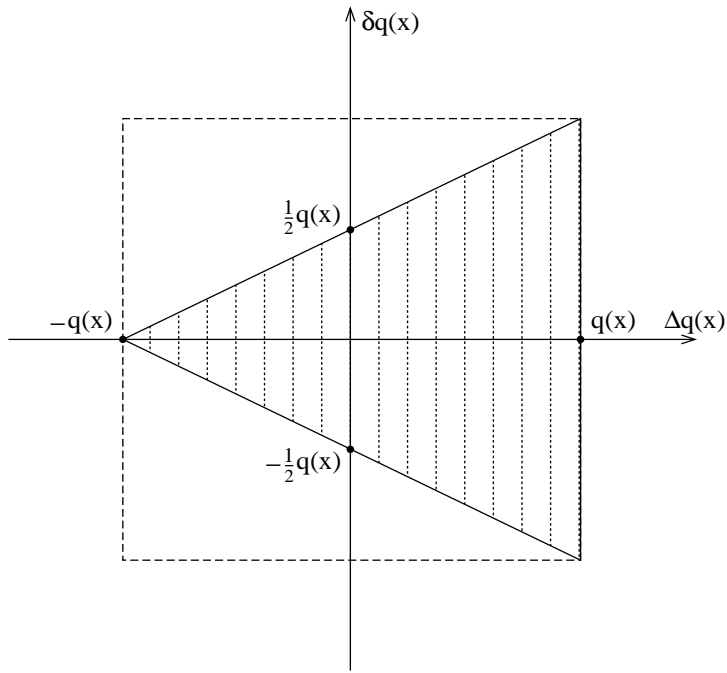,width=10cm,height=10cm}}
\vspace*{-1cm}
\caption[]{\small The hatched area represents the domain allowed by 
positivity, Equation~\ref{tin2}.}
\label{ffig1} 
\end{figure}

The joint description of the quark distributions in terms of the
${\cal A} (H,h;H',h')$ implies that transversity is not entirely
unrelated to the $q$,$\Delta q$. Indeed, rewriting~\cite{goldstein95}
${\cal A}(H,h;H',h')=\sum_X a^*_{H'h'} (X) a_{Hh}(X)$, 
where $X$ is an arbitrary final state, one finds from the condition 
$\sum_X |a_{++}(X) \pm a_{--}(X)|^2 \geq 0$ the inequality~\cite{soffer95}
\begin{equation}
q(x) + \Delta q(x) \geq 2|\delta q(x)| \, \, . \label{tin2}
\end{equation}
 Figure~\ref{ffig1} displays the region allowed by Equation~\ref{tin2}, which 
is indeed smaller than the one resulting from the trivial condition 
$|\delta q(x)|\leq q(x)$. Equation~\ref{tin2} holds for all quark 
flavors and separately for their corresponding antiquarks. As was 
demonstrated in References~\cite{barone97,vogelsang98,bourrely98,omartin98},
the inequality is preserved under QCD evolution; that is, if it is assumed 
to be satisfied at one resolution scale, it will hold at all larger scales.
This remains true~\cite{vogelsang98,bourrely98,omartin98} even at 
two-loop order~\cite{vogelsang98,koike97} in evolution. 

The helicity flip required for transversity to contribute to hard
scattering can occur if there are two soft hadronic 
vertices in the process. In this case, transverse spin can be 
carried from one hadron to the other along a quark line.
One possibility is to have 
two transversely polarized hadrons 
in the initial state, as realized at RHIC. 
An alternative is to have one transversely polarized initial hadron and a final-state 
fragmentation process that is sensitive to transverse polarization. Here, 
the other initial particle could be a lepton, as in DIS, or another
proton, as 
at RHIC. 

For 
the first possibility, a promising candidate process for a measurement 
of the $\delta q$, $\delta \bar{q}$ is Drell-Yan dimuon production 
which, to lowest order in QCD, proceeds via $q\bar{q}\rightarrow 
\gamma^{*}$ annihilation. A systematic study of this process was in 
fact also the place where the transversity densities made their first 
appearance in theory~\cite{ralston79}. On the downside of this reaction 
is that the transversity antiquark density in the nucleon is 
presumably rather small; there is no splitting term $g\rightarrow 
q\bar{q}$ in the evolution equations for transversity~\cite{artru90}, 
so a vital source for the generation of antiquarks is missing (only 
higher orders in evolution produce antiquarks carrying 
transversity~\cite{vogelsang98,omartin98}). Also, in Drell-Yan, the event 
rate is generally low. However, when compared to other conceivable 
reactions in $pp$ collisions that serve to determine parton densities, 
the Drell-Yan process has the advantage that to lowest order there is 
no partonic subprocess that involves a gluon in the initial state. If a 
reaction does have a gluon-initiated subprocess, its transverse
double-spin asymmetry is expected to be suppressed~\cite{xji92,jaffe96}. 
This is 
because gluons usually strongly contribute to the unpolarized cross sections 
in the denominator of the asymmetry, whereas they are absent for
transversity, as discussed above. In addition, for many reactions other
than Drell-Yan, one finds a particular ``selection-rule''~\cite{xji92,jaffe96}
suppression of the contributing transverse subprocess asymmetries.

Several phenomenological studies of Drell-Yan dimuon production at RHIC 
have been 
presented~\cite{vogelsang93,contogouris94,bourrely95,calarco97,omartin98,miyama99}. Model estimates of the transversity densities have been obtained 
in these studies by either assuming $ 2\delta q(x,Q^2_0)=q(x,Q^2_0) + 
\Delta q(x, Q^2_0)$ (see Equation~\ref{tin2}), or by 
employing~\cite{miyama99,scopetta98} $\delta q(x,Q^2_0) \simeq 
\Delta q(x, Q^2_0)$, at some initial (low) resolution scale $Q_0$.
%
%
Note that the latter ansatz violates 
inequality~\ref{tin2} if
$\Delta q(x,Q^2_0)<-\frac{1}{3} q(x,Q^2_0)$. The transverse double-spin 
asymmetry for Drell-Yan dimuon production is (to lowest order) 
\begin{equation} \label{att}
A_{TT} = \widehat a_{TT} \frac{\sum_{q}^{} e^2_q \delta q
(x_1,M^2) \delta \bar{q} (x_2,M^2) + (1 \leftrightarrow 2)}
{\sum_{q}^{} e^2_q  q (x_1,M^2) \bar q (x_2,M^2) + 
(1 \leftrightarrow 2)}.
\end{equation}
Here $\widehat a_{TT}$ is the partonic transverse-spin asymmetry, 
calculable in perturbative QCD, and $M$ is the dilepton mass.
NLO corrections to Drell-Yan dimuon production with transversely polarized 
beams have 
been calculated~\cite{vogelsang93,contogouris94,omartin98,contog98,kamal96} 
and are routinely used in numerical studies.

\begin{figure}[ht]
\centerline{\psfig{figure=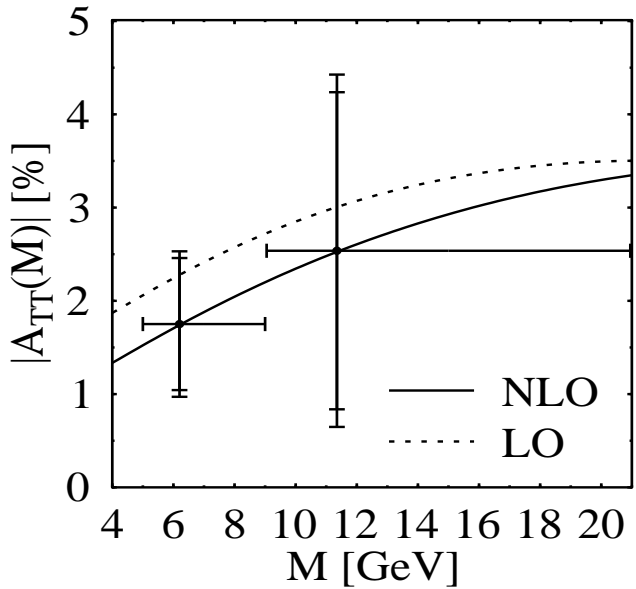,width=10cm,height=9cm}}
\caption[]{\small 
Next-to-leading-order transverse-spin asymmetry for Drell-Yan 
dimuon production at $\sqrt s =200$~GeV~\cite{omartin98}.} 
\label{cr5} 
\end{figure}

The {\sc Phenix} endcaps will be able to identify muons with rapidity
$1.2<|y_{\mu^{\pm}}| <2.4$. Figure~\ref{cr5} shows 
predictions~\cite{omartin98} for $A_{TT}$. In order to model the transversity 
densities, saturation of inequality~\ref{tin2} at a low scale 
$Q\approx 0.6$ GeV has been assumed, making use of the information on the 
$\Delta q$, $\Delta \bar{q}$ 
in that inequality coming from polarized DIS.
The statistical errors expected for {\sc Phenix} are 
also shown. One observes that the asymmetry is generally small but 
could be visible experimentally if the transversity densities are not 
much smaller than those used here. Larger estimates for $A_{TT}$ 
have been obtained 
\cite{bourrely95}, based 
on more optimistic assumptions concerning the size of the $\delta q$, 
$\delta \bar{q}$. 
Careful studies of the background to 
lepton pair production resulting from coincidental semileptonic 
heavy-flavor decays (see Section~\ref{wasy}) will be important.

The other possibility involves one transversely polarized initial hadron and a final-state 
fragmentation process that is sensitive to transverse polarization. Promising approaches have emerged from considering 
the production of high-$p_T$ dimeson systems~\cite{collins94,ji94,jjt97}, or
from taking into account ``intrinsic'' transverse momentum degrees of freedom 
in a fragmentation process producing a single high-$p_T$ pion~\cite{collins93}.
Both dimesons and pions are very abundantly produced in high-energy $pp$ 
collisions. It has been shown~\cite{collins94} that the azimuthal 
distribution of low-mass pairs of pions about the final-state jet axis can
be used as a measure of the transverse polarization of the quark initiating 
the jet. The same is true~\cite{collins93} for the ``intrinsic'' transverse 
momentum distribution of a produced pion relative to its quark progenitor. 
In this way, one effectively obtains an asymmetry that is sensitive to 
products of the transversity density for the initial-state quark and a
transverse-polarization--dependent fragmentation function for 
the final state. For instance, for the mechanism proposed for DIS
in Reference~\cite{collins93}, the fragmentation function would be
\begin{equation} \label{collfct}
H_1^{\perp} (z,k_{\perp}) \, \propto \, 
D_{\pi/\qup}(z,k_{\perp}) -
D_{\pi/\qdown}(z,k_{\perp})\;,
\end{equation}
where $k_{\perp}$ is the ``intrinsic'' transverse momentum in the
fragmentation process. Notice that one polarized proton in the initial 
state is sufficient for this kind of measurement. Time-reversal 
invariance, however, precludes a nonzero effect unless phases are 
generated by final-state interactions in the fragmentation process that 
do not average to zero upon summation over unobserved hadrons. It is 
a~priori unclear whether such a net phase will exist. This 
led to investigation~\cite{jjt97} 
of the interference between $s$ and $p$ 
waves of two-pion systems with invariant mass around
the $\rho$. Such an interference effect yields sensitivity to the 
polarization of the quark progenitor through the quantity 
$\vec{k}_{\pi^+} \times \vec{k}_{\pi^-} \cdot \vec{s}_T$, where the 
$\vec{k}$s are the pion momenta and $\vec{s}_T$ is the transverse 
nucleon spin; one effectively uses the angular momentum of the two-pion 
system as a probe of the quark's polarization. Staying in the mass region 
around the $\rho$ ensures~\cite{jjt97} that the final-state interaction 
phase does not average to zero. The $s$-$p$ wave interference in the
$q\rightarrow \pi\pi$ formation is described by a new set of fragmentation 
functions, the interference fragmentation functions~\cite{jjt97}. 
Just as the function in Equation~\ref{collfct}, the latter are presently 
entirely unknown; the price to be paid for obtaining sensitivity to 
transversity in all of the ways suggested in Reference~\cite{collins94,jjt97,collins93} 
is thus the introduction of another unknown component. However, one may hope 
that the involved fragmentation functions can be determined independently in 
$e^+ e^-$ annihilation. Studies of the experimental situation at RHIC 
concerning the proposal of~\cite{jjt97} are under way~\cite{matthias}.
\subsection{Transverse Single-Spin Asymmetries}\label{subsec:sspin}
\begin{figure}[h]
\begin{center}
\centerline{\psfig{figure=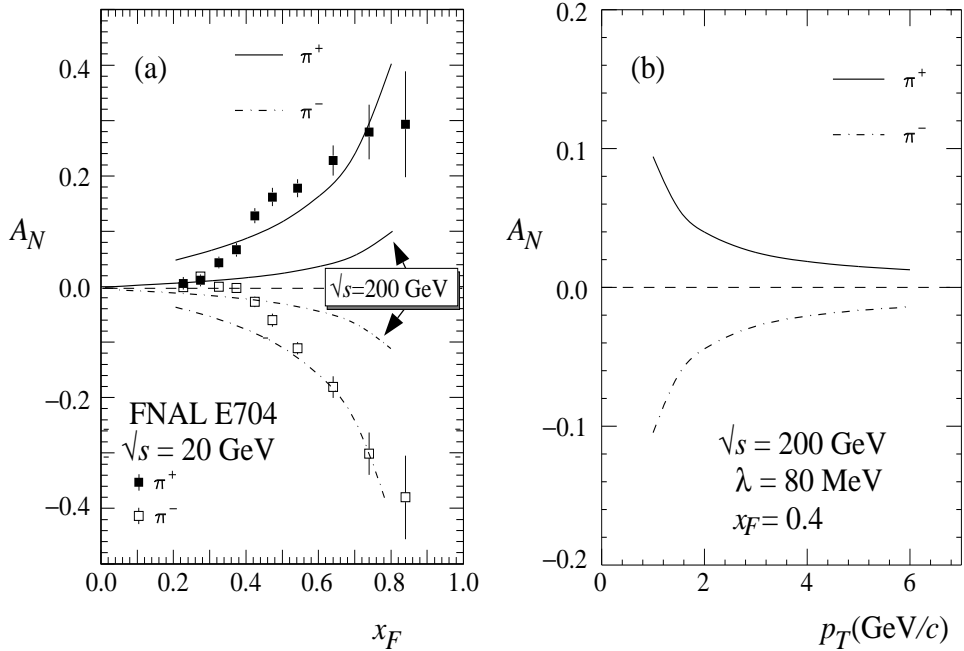,width=13cm,height=9cm}}
\caption[]{\small ($a$) Experimental data~\cite{adams91} and 
theoretical calculations~\cite{qiu98} for transverse single-spin 
asymmetries for $\pi^+$ and $\pi^-$ production in $pp$ collisions at 
$\sqrt{s}$=20~GeV as functions of $x_F$. Predictions for 
RHIC for $p_T$=4~GeV/$c$ are superimposed. The 
transverse momentum 
dependence for RHIC at $x_F=0.4$ is shown in ($b$).}
\label{pion} 
\end{center}
\end{figure}

Surprisingly large single-transverse spin asymmetries, for instance in
fixed-target $p^{\uparrow} p\rightarrow \pi X$ at pion transverse 
momenta of a few GeV, have been observed 
experimentally~\cite{bunce76} over many years. RHIC will further 
investigate the origin of such asymmetries. Within the 
``normal'' framework of perturbative QCD and the factorization theorem 
at twist-2 for collinear massless parton configurations, no 
single-transverse spin asymmetry is obtained---nonzero effects 
occur only when 
one keeps quark mass terms (as is required to generate 
helicity flips) and when 
one takes into account at the same time 
higher-order loop diagrams that produce relative 
phases~\cite{pumplin78}. Such effects are therefore of the order of
$\alpha_s m_q/\sqrt{s}$ and 
cannot explain data such 
as that in Reference~\cite{bunce76}. It is believed that nontrivial higher-twist effects
are responsible for the observed single-spin 
asymmetries~\cite{efremov85,qiu91,qiu98}. Reference~\cite{qiu91,qiu98} 
showed how single transverse-spin asymmetries can be evaluated 
consistently in terms of a generalized factorization theorem
in perturbative QCD, wherein they arise, for example, as convolutions 
of hard-scattering functions with an ordinary twist-2 parton density 
from the unpolarized hadron and a twist-3 quark-gluon correlation function 
representing the polarized hadron. Another contribution involves
the transversity distribution and another (chiral-odd) spin-independent 
twist-3 function of the proton~\cite{qiu98,koike00}. A simple model was 
constructed~\cite{qiu91,qiu98} that assumes only correlations of valence 
quarks and soft gluons. It can describe the present data and makes 
various definite predictions, to be tested at RHIC, where one certainly 
expects to be in the perturbative domain. In particular, at RHIC, one
should see the fall-off with $p_T$ of the single-transverse spin asymmetries 
in single-inclusive pion production,
associated with their twist-3 nature (see Figure~\ref{pion}).

A related dynamical origin for transverse single-spin asymmetries was 
proposed~\cite{sivers90,collins93,boer98} to reside in the dependences of 
parton distribution and fragmentation functions on intrinsic parton 
transverse momentum $k_T$. In fact, the proposal of~\cite{collins93} 
for measuring transversity in the proton, which we discussed in the 
previous subsection, proceeds for $pp$ scattering exactly through a 
single-transverse spin asymmetry, making use of the $k_T$-dependent 
fragmentation function in Equation~\ref{collfct}. Suppression of the asymmetry 
should also arise here, through a factor $\langle k_T \rangle/p_T$.
It has also been considered that single-spin asymmetries might be 
generated by $k_T$ dependences of the parton distribution functions 
in the initial state~\cite{sivers90,boer98}. Here, 
one could have
\begin{eqnarray} \label{fcts}
f_{1T}^{\perp}(x, k_{\perp}) &=&
f_{q/\pup}(x, k_{\perp}) -
f_{q/\pdown}(x, k_{\perp})\;,  \nonumber \\
h_1^{\perp}(x, k_{\perp})  &=&
f_{\qup/p}(x, k_{\perp}) -
f_{\qdown/p}(x, k_{\perp})\; 
\end{eqnarray}
as the driving forces. There is a qualitative difference between the 
functions in  Equations~\ref{fcts} and \ref{collfct}: In order to be able to 
produce an effect, the latter requires final-state interactions 
(which are certainly present), to make the overall process 
time-reversal-symmetry-conserving (see the previous subsection). In contrast, the 
distributions in Equation~\ref{fcts} rely on the presence of nontrivial 
(factorization-breaking) initial-state interactions between the incoming 
hadrons~\cite{anselmino95}, or on finite-size effects for the 
hadrons~\cite{sivers98}; they vanish if the initial hadrons are described
by plane waves. This makes the ``Collins function'' (Equation~\ref{collfct}) perhaps 
a more likely source for single-spin asymmetries. The reservations
concerning Equation~\ref{fcts} notwithstanding, when 
a factorized 
hard-scattering model is evoked, each mechanism described by 
Equations~\ref{collfct} and \ref{fcts} can by itself account 
for~\cite{anselmino95,boer98,leaderb99} the present $p^{\uparrow} p
\rightarrow \pi X$ data. Also, all could be at work simultaneously 
and compete with one another. Single-spin Drell-Yan measurements at 
RHIC should be a good testing ground~\cite{boer98} for the existence of 
effects related to Equation~\ref{fcts}, since for Drell-Yan the 
Collins function (Equation~\ref{collfct}) cannot contribute.
\subsection{Spin-Dependent Fragmentation Functions}
Even in the context of a parity-conserving theory like QCD, an asymmetry 
can arise for only one longitudinally polarized particle in the 
initial state, if the longitudinal polarization of a particle in the 
final state is observed. 
The measurement of the polarization of an outgoing highly energetic 
particle certainly provides a challenge to experiment. 
$\Lambda$ baryons are particularly suited for such studies, thanks to the 
self-analyzing properties of their dominant weak decay, $\Lambda \rightarrow 
p \pi^-$. Recent results on $\Lambda$ production reported from 
LEP~\cite{buskulic96} have demonstrated the feasibility of 
successfully reconstructing the $\Lambda$ polarization.

\begin{figure}[h]
\centerline{\psfig{figure=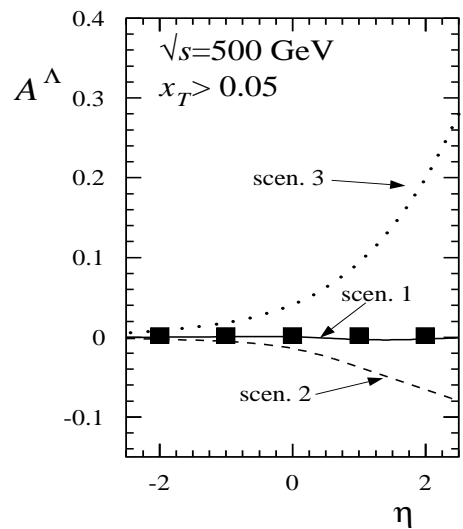,width=6.cm,height=7.0cm}}
\caption{\small The longitudinal spin transfer asymmetry in $\Lambda$ 
production at RHIC ($\protect\sqrt{s}=500$~GeV)~\protect\cite{deflorian98p}, 
as a function of rapidity of the $\Lambda$
for various sets of spin-dependent fragmentation functions proposed
in Reference~\protect\cite{deflorian98l}. For scenario~1,
only strange quarks transmit polarization to the $\Lambda$. In scenario~2, 
there is also a (negative) contribution from up and down 
quarks~\protect\cite{burkardt93}. In scenario~3, all quarks equally produce 
polarized $\Lambda$s. 
The expected errors for {\sc Star} with standard luminosity and polarization
are comparable to 
the ``data'' shown for $\eta =\pm 2$, and smaller for the other points.}
\label{lam}
\end{figure}

Spin-transfer asymmetries give information on yet unexplored 
spin effects in the fragmentation process. For our $\Lambda$ example, 
the longitudinal transfer asymmetry will be sensitive to the
functions
\begin{equation}
\label{eq:ffdef}
\Delta D_i^{\Lambda}(z) \equiv D_{i(+)}^{\Lambda (+)}(z) -
 D_{i(+)}^{\Lambda (-)}(z)
\end{equation}
describing the fragmentation of a longitudinally polarized parton 
$i=q,\bar{q},g$ into a longitudinally polarized $\Lambda$, where 
$D_{i(+)}^{\Lambda (+)}(z)$ $(D_{i(+)}^{\Lambda (-)}(z))$ is the probability 
of finding a $\Lambda$ with positive (negative) helicity in a parton 
$i$ with positive helicity, carrying a fraction $z$ of the parent parton's
momentum (see Section~\ref{sec:PREREQ}). As was shown 
in Reference~\cite{deflorian98l,ma99}, 
the LEP measurements~\cite{buskulic96} have provided 
initial information 
on some combinations of the $\Delta D_i^{\Lambda}$ but leave room for very 
different pictures of the spin-dependence in $\Lambda$ fragmentation. 
Measurements of the polarization of $\Lambda$s produced in $\vec{p}p$ 
collisions at RHIC should vastly improve~\cite{deflorian98p,boros00} our 
knowledge of the $\Delta D_i^{\Lambda}$. Figure~\ref{lam} illustrates 
this by showing the longitudinal spin transfer asymmetry at RHIC, 
defined in analogy with Equation~\ref{allsigma} as 
\begin{equation}
A^{\Lambda}=\frac{(\sigma^{\Lambda(+)}_{+}+\sigma^{\Lambda(-)}_{-})-
(\sigma^{\Lambda(+)}_{-}+\sigma^{\Lambda(-)}_{+})}
{(\sigma^{\Lambda(+)}_{+}+\sigma^{\Lambda(-)}_{-})+
(\sigma^{\Lambda(+)}_{-}+\sigma^{\Lambda(-)}_{+})},
\end{equation}
where the lower helicity index refers to the polarized proton and the upper
to the produced $\Lambda$. Various models for the $\Delta D_i^{\Lambda}$, 
all compatible with the LEP data, have been used in Figure~\ref{lam}.
It will be interesting to see which scenario is favored by the RHIC 
measurements. A cut of $x_T>0.05$ has been applied in the figure. 
$\Lambda$s are very copiously produced at RHIC~\cite{deflorian98p}, resulting
in small expected statistical errors.

Similarly optimistic 
conclusions have been reached~\cite{deflorian98s} for the case of 
transverse polarization of one initial beam and the $\Lambda$, 
in which case RHIC experiments would yield information on the product
of the proton's transversity densities and the transversity fragmentation
functions of the $\Lambda$, which are both so far unknown.

\clearpage
\section{PHYSICS BEYOND THE STANDARD MODEL}

So far we have discussed probing the proton spin structure at RHIC, and both using and testing
perturbative QCD in the spin sector.  Spin is also an excellent tool
to go beyond the standard model and to uncover important new physics, if it exists. 
Many extensions of the standard model have 
been proposed. Our purpose in this section is to illustrate
this new potentiality by means of a specific example.
 
\begin{figure}[ht]
\centerline{\psfig{figure=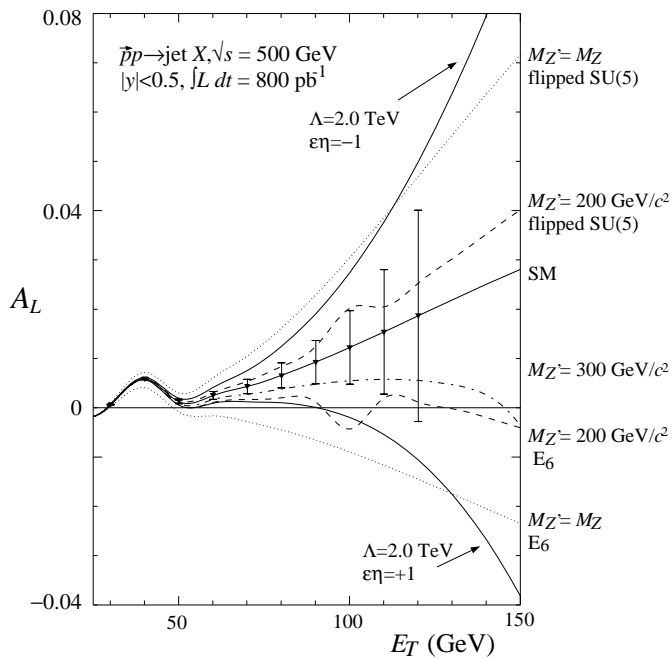,width=9cm}}
\caption{\small $A_{L}$ for one-jet inclusive production in 
$\vec{p}p$ collisions, versus 
transverse energy, for $\protect\sqrt s=500$~GeV. 
The solid curve with error bars represents the standard-model expectation.
The error bars show the sensitivity at RHIC for 
800~pb$^{-1}$, for the {\sc Star} detector.
The other solid curves, labeled by the product of $\epsilon\eta$,
correspond to the contact interaction at $\Lambda = 
2$~TeV~\protect\cite{TVCI}.
The dashed and dotted curves correspond to 
different leptophobic $Z'$ models \protect\cite{TVZp}.
The calculations are at leading order.}
\label{fig:alby}
\end{figure}

Let us consider one-jet inclusive production. As discussed in Section~3,
the cross section is dominated by the pure QCD $gg$, $gq$, and
$qq$ scatterings, but the existence of the electroweak interaction,
via the effects of the $W^{\pm}$ and $Z$ gauge bosons, adds a small contribution. 
Consequently, the parity-violating helicity asymmetry $A_L$, defined as~\cite{Tan91}
\begin{equation}
A_L = -\left [ {{d\sigma}_{+}^{pp \rightarrow jet X}\over{dE_T}}-{{d\sigma}_{-}^{pp \rightarrow jet X}\over{dE_T}} 
\right ]\cdot 
\left [ {{d\sigma}_{+}^{pp \rightarrow jet X}\over{dE_T}}+{{d\sigma}_{-}^{pp \rightarrow jet X}\over{dE_T}} \right ]^{-1},
\end{equation}
is expected to be nonzero from the QCD-electroweak interference (as shown
in Figure~\ref{fig:alby}). Additionally, a small peak near $E_T=M_{W,Z}/2$ 
is seen, which is the main signature of the purely electroweak contribution.
The cross sections are for one longitudinally polarized beam,
colliding with an unpolarized beam.  The existence of new parity-violating 
interactions could lead to large modifications of this standard-model 
prediction~\cite{Tan91}.

First let us recall that the sensitivity to the presence of some new quark-quark contact interactions 
has been analyzed in Reference~\cite{TVCI}. Such a contact interaction could represent the effects of quark compositeness, under the form
\begin{equation}
{\cal L}_{qqqq} = \epsilon \, {g^2\over {8 \Lambda^2}} 
\, \bar \Psi \gamma_\mu (1 - \eta \gamma_5) \Psi \cdot \bar \Psi
\gamma^\mu (1 - \eta \gamma_5) \Psi~,
\end{equation}
where $\Psi$ is a quark doublet, $\Lambda$ is a compositeness scale, and 
$\epsilon=\pm 1$. If parity is maximally violated,
$\eta=\pm 1$. Figure~\ref{fig:alby} shows how the standard-model prediction 
will be affected by such a new interaction, assuming $\Lambda=2$~TeV, 
which is close to                                  the present limit obtained 
for example by the {{\sc DO}\hspace{-0.7em}/}  experiment at the Tevatron~\cite{D0}. 
The statistical errors shown are for standard RHIC luminosity of 800~pb$^{-1}$, and for jets with rapidity
$|y|<$0.5, and include measuring $A_L$ using each beam, summing over the spin states of the
other beam.  Due to the parity-violating signal's sensitivity to new physics, RHIC is surprisingly
sensitive to quark substructure at the 2-TeV scale and is competitive with the Tevatron, 
despite the different energy ranges of these machines. Indeed, a
 parity-violating signal beyond the standard model at RHIC would
definitively indicate the presence of new physics~\cite{Tan91}.

RHIC-Spin would also be sensitive to possible new neutral
gauge bosons \cite{TVZp}. A class of models, called leptophobic $Z'$,
is poorly constrained up to now. Such models appear naturally in several
string-derived models \cite{leptoZ} (nonsupersymmetric 
models may be also constructed \cite{Agashe}). In addition,
in the framework of supersymmetric models with an additional Abelian
gauge factor $U(1)'$, it has been shown \cite{CL} that the $Z'$ boson could appear
with a relatively low mass ($M_Z \le M_{Z'} \le$1~TeV) and
a mixing angle with the standard $Z$ close to zero.
The effects of different representative models
are shown in Figure~\ref{fig:alby} (see Reference~\cite{TVZp} for details).
RHIC
covers some regions in the parameter space
of the different models that are unconstrained by present and forthcoming experiments
(e.g.\ Tevatron Run~II), and
 RHIC would also uniquely obtain
information on the chiral structure of the new interaction.\\

Other possible signatures of new physics at RHIC have been investigated. 
Particularly interesting quantities~\cite{BOURRELY93,Ryk,KSS} are transverse 
(single or double) spin asymmetries for $W^{\pm}$ production, since these 
are expected to be extremely small in the standard 
model~\cite{BOURRELY93,dboer00}. 
For instance, the case of the corresponding standard-model double spin 
asymmetry $A_{TT}^{\pm}$ was examined in detail recently~\cite{dboer00}. 
Non-vanishing contributions could arise here for example in the form of 
higher-twist terms, which would be suppressed as powers of $M^2/M_W^2$, 
where $M$ is a hadronic mass scale and $M_W$ the $W$ mass. Other possible 
contributions were demonstrated in~\cite{dboer00} to be negligible as well. 
By similar arguments, also the corresponding single-transverse spin asymmetry 
for $W^{\pm}$ production, $A_N^{\pm}$, is expected to be extremely small in 
the standard model~\cite{KSS}. New physics effects, on the contrary, might 
generate asymmetries at leading twist, for example through non-$(V-A)$ 
(axial)vector couplings of quarks to the $W$, or through tensor or (pseudo)scalar 
couplings, all of which would also have to violate ${\rm CP}$ in order to generate
a single-spin asymmetry $A_N^{\pm}$. In particular the latter asymmetry has been 
examined with respect to sensitivity to new physics effects at RHIC~\cite{KSS}. 
For a case study, the minimal supersymmetric extension of the standard 
model, with ${\rm R}$-parity violation, was employed, which contains scalar 
quark-$W$ interactions and complex phases, resulting in ${\rm CP}$-violating 
effects. The results of~\cite{KSS} show that in this particular extension of
the standard model, $A_N^{\pm}$ is likely to be very small as well, 
below the detection limit of RHIC. Nevertheless, this
does not exclude that other non-standard mechanisms produce larger effects,
and $A_N^{\pm}$ and $A_{TT}^{\pm}$ will be measured at RHIC with transversely
polarized beams in the context of the physics discussed in the previous
section.  A non-zero result would be a direct indication of new physics.

\clearpage
\section{SMALL-ANGLE $pp$ ELASTIC SCATTERING}
In previous sections, we have discussed the physics of hard scattering at RHIC with
polarized protons, which can be understood as collisions of polarized quarks
and gluons.  The scattering is so energetic that we can use perturbative QCD to
describe the interactions of the quarks and gluons, and, thus, probe the spin
structure of the proton at very small distances.  For example, scattering at
Q$^2$=(80~GeV)$^2$ probes wave lengths of 0.003~fermi.  Small-angle scattering,
from total cross section to $t=-1$~(GeV/$c$)$^2$, probes the static proton properties
and constituent quark structure of the proton, covering distances from 4~fermi
[$-t=0.003$~(GeV/$c$)$^2$ in the Coulomb nuclear interference (CNI) region] to a distance of $\approx$0.2~fermi.  
Unpolarized scattering shows striking behavior in this region, from the surprise
that total cross sections rise at high energy, to observed dips in elastic cross
sections around $-t=1$~(GeV/$c$)$^2$.  The {\sc pp2pp} experiment at RHIC~\cite{Guryn} will explore this
region for spin-dependent cross sections, for $\sqrt{s}$=20-500~GeV, for the first time.

Historically, new spin-dependent data have often shown new structure underlying
spin-independent cross sections, indicating the presence of unexpected dynamics in
the interaction.  Several examples have been discussed in previous sections.  
Previous work with spin stops at $\sqrt{s}$=20~GeV,
where tertiary polarized $p$ and $\bar{p}$ beams were collected from the 
parity-violating
decays of $\Lambda$ and $\bar{\Lambda}$ hyperons and steered onto unpolarized hydrogen
and polarized pentanol ($C_5H_{12}O$) targets~\cite{E704}.  RHIC will provide much
higher intensity, a large extension of the energy range, and pure targets for
2-spin measurements.  

In the energy regime $\sqrt{s}>$20~GeV,
total cross sections have been observed to rise with energy for
$pp$, $\bar{p}p$, $\pi^{\pm}p$, and $K^{\pm}p$.  The $\bar{p}p$ total cross section
rises through the Tevatron maximum energy of 2~TeV, and the $pp$ total cross section
has been observed to rise through its highest energy measurement at the ISR,
$\sqrt{s}$=62~GeV~\cite{totalsigma}.  The {\sc pp2pp} experiment will
measure spin-dependent total cross sections, $\sigma_{\uparrow\uparrow}$, 
$\sigma_{\uparrow\downarrow}$, and $\sigma_L = \sigma_+ - \sigma_-$ [where the arrows
represent transverse spin measurements, and (+) and ($-$) represent helicities] through
the range of rising cross sections available at RHIC.  The unpolarized $pp$ total 
cross section measurements will also be extended to $\sqrt{s}$=500~GeV.    

For $\bar{p}p$, the rise of the total cross section has 
been successfully described in the impact picture approach  on the basis of the high-energy 
behavior of a relativistic quantum field theory \cite {CH70}. 
This is based on the fact that the effective interaction strength increases with energy
in the form $s^{1+c}/(ln~s)^{c'}$, a simple expression in two key parameters $c$ and $c'$,
where $s$ is expressed in GeV$^2$.
A fit of the data then leads to the values of the two free parameters $c=0.167$, $c'=0.748$ \cite{BSW84}. 
If this picture is correct (the field theoretical argument is based on connecting QED and QCD
theories, but successfully predicted that the $\bar{p}p$ total cross section would continue to
rise, following these parameters), there should be no difference in the rise of $pp$ and $\bar{p}p$
total cross sections.  An extension of this approach allows a description of the elastic
cross section~\cite{others}, which will also be measured at RHIC.

The single-spin asymmetry for $pp$ elastic scattering, $A_N$, is expected to be small 
but significant in the CNI region, 
$-t=0.001$--$0.01$~(GeV/$c$)$^2$~\cite{BKLST}.  As discussed previously,
$pp$ elastic scattering in the CNI region will be the basis of the RHIC
polarimetry.  CNI scattering is expected to produce an asymmetry from scattering
an unpolarized proton (polarization averaged to zero) in one beam from
the magnetic moment of a polarized proton from the other beam, with a maximum of
$A_N=0.04$ at $-t=0.003$~(GeV/$c$)$^2$.  However, a hadronic spin-flip term can also
contribute to the maximum, and this term is sensitive to the static constituent
quark structure of the proton.  The authors of Reference~\cite{BKLST} remark that the helicity 
flip probes the shortest interquark distance in the
proton, and that the helicity nonflip is sensitive to the largest quark
separation in the proton due to color screening.  The helicity-flip term, if present,
can indicate an isoscalar anomalous
magnetic moment of the nucleons~\cite{bourrely75}, an anomalous color-magnetic moment causing
helicity nonconservation at the constituent quark-gluon vertex~\cite{ryskin},
and/or a compact quark pair 
%
%
in the proton~\cite{kz,zak}.

\begin{figure}[h]
\centerline{\epsfbox{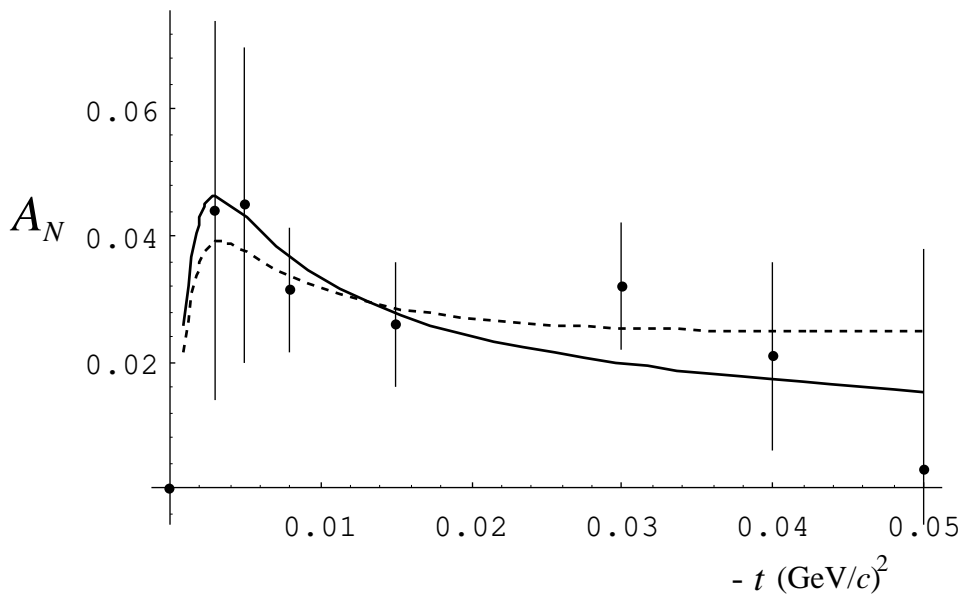}}
\medskip
\caption{\small Transverse single-spin asymmetry for 
proton proton elastic scattering.
The data points are from Fermilab E704 \protect\cite{E704}. The solid curve
is the best fit with the spin-flip hadronic amplitude
constrained to be in phase with non-flip hadronic amplitude;
the dotted curve is the best fit without this constraint.}
\label{fig:ane704}
\end{figure}

The only measurement of $A_N$ in the CNI region at higher energy is by E704 at Fermilab~\cite{E704} 
at a lab
momentum $p_L=200 \mbox{ GeV}/c$; the results are shown in Figure~\ref{fig:ane704}.
The errors are too large to allow an unambiguous theoretical
interpretation.  There are two fits to the E704 data shown with a nonzero hadronic 
spin flip term~\cite{trueman}.
As emphasized in References~\cite{trueman} and \cite{bs},
a large value of the hadronic helicity-flip amplitude
generates a very large change in the maximum in $A_N$, which
can be of the order of 30\% or more. 
The {\sc pp2pp} experiment will measure $A_N$ to $\pm$0.001 in the CNI peak.
This level of precision is required for absolute polarimetry, giving an
expected precision on $A_N$ of $\Delta A_N/A_N$=$\pm$0.001/0.04=$\pm$0.025.
This experiment will cover from 
$0.0005 \le -t \le 1.5$~(GeV/$c$)$^2$ (with additional
detectors for the larger $-t$ region).  Thus, the location of the maximum in $A_N$
and its maximum value and shape will be determined.

Small-angle scattering at high energy is presently understood in the Regge picture
as being dominated by Pomeron exchange~\cite{pdbcollins}.  The Pomeron, which has the vacuum quantum numbers
with charge-conjugation $C=+1$, can be interpreted as a two-gluon exchange.
There is room in the data for a small three-gluon exchange contribution with 
$C=-1$, the Odderon~\cite{Odderon}.  It has been shown recently~\cite{LT00} that the behavior of
the two-spin transverse asymmetry $A_{NN}$ in $pp$ elastic scattering in the CNI region depends
strongly on the Odderon contribution and that the {\sc pp2pp} experiment is quite sensitive to
its presence.

In addition to the measurements discussed above, the {\sc pp2pp} experiment will measure 
larger angles, to $-t=1.5$~(GeV/$c$)$^2$, which includes the region of dip structure in
the unpolarized cross section, measuring $A_N$ and the two-spin asymmetries $A_{NN}$,
$A_{SS}$, and $A_{LL}$~\cite{BKLST}. 
These first, precise, determinations of spin dependence for small-angle $pp$ elastic
scattering in the energy
range 
$\sqrt{s} = 20$--500~GeV probe the spin
structure of the proton in the nonperturbative QCD region, from the static properties 
of the proton to its constituent quark structure. 

At higher energy, such as at the LHC, the CNI region becomes inaccessible.
The minimum $-t$ reachable with colliding beams depends on scattering the protons out of
the beams.  For fixed $-t$, the scattering angle falls as 1/p$_{{\rm beam}}$, whereas the beam
size falls more slowly as 1/$\sqrt{{\rm p}_{beam}}$.  Roughly, this limits an experiment
at the LHC to $-t>$0.01~(GeV/$c$)$^2$.

\section{CONCLUDING REMARKS}

RHIC will 
be the first machine to look at the proton spin structure by colliding polarized proton 
beams rather 
than scattering polarized leptons off polarized targets.  Thus, one can test fundamental 
interactions 
in an entirely different environment and at much higher energies, 
as in the unpolarized case. (Here, too, information on the nucleon structure from DIS has been complemented 
by information from hadron colliders.)  
For hadron colliders, including 
RHIC-Spin, due to the high energy and luminosity 
that give access to hard parton 
scattering, perturbative QCD
probes in one proton are used to study the nonperturbative structure of the 
``target'' proton. 

What can we expect 
from 
RHIC-Spin? If, for example, a large gluon polarization is observed, such 
a signal would imply 
a previously unknown fundamental role of the gluons in the proton spin.  Surprise and new 
insights are very likely.  

This field is very new both theoretically and experimentally.  Previous experimental 
spin work with hadron probes was at much lower energy and luminosity, and used 
impure polarized targets.  Much of the discussion presented here is from very 
recent work.  Thus, this article should not be seen as a review but rather 
as an invitation.

\section*{ACKNOWLEDGMENTS}
This article summarizes the hard work of many collaborators in theoretical, experimental, and accelerator physics
who have developed
the RHIC spin program.
Leaders in this program from its beginnings
to the present include Michael Tannenbaum, Yousef Makdisi, and Thomas Roser 
of Brookhaven;
the Marseille Theory Group; and Aki Yokosawa, Hal Spinka, and Dave Underwood 
of Argonne. The Kyoto, Penn State, UCLA, and IHEP groups have 
contributed many ideas and
studies, including Vladimir Rykov of Wayne State University (previously of IHEP).  
Robert Jaffe of MIT helped initiate a close collaboration
between theorists and experimenters to develop the RHIC spin program.
We acknowledge important ideas and work by Leslie Bland and colleagues at
Indiana University.  
In addition, we would like to thank Daniel Boer, 
Akira Masaike, Marco Stratmann,  
Lawrence Trueman, Jean-Marc Virey.
The key
step in this spin program, leading to its anticipated first run in 2001,  
has been the involvement of RIKEN, Japan. RIKEN has supported, beginning 
in 1995, the spin hardware including the
Siberian Snakes and Spin Rotators, and a second muon arm for spin for 
{\sc Phenix}; it created the
RIKEN BNL Research Center to develop our understanding of RHIC
physics, and it supports a strong spin group based at Brookhaven.  
The Science and Technology Agency of Japan supports RIKEN.
We thank the US Department of Energy for its early support of the RHIC spin program.

\section*{Appendix: Information from Polarized Deep-Inelastic Scattering}
In this Appendix we briefly discuss the information from DIS 
on $\Delta q, \Delta \bar{q},\Delta g$. 
If we neglect contributions resulting from $W^{\pm}$ or $Z^0$
exchange, DIS is sensitive only to the sums of quarks and antiquarks 
for each flavor. Therefore, we define
\begin{equation}
\Delta {\cal Q} (x,Q^2) \equiv \Delta q  (x,Q^2) + \Delta \bar{q} (x,Q^2) \, .
\end{equation}
To lowest order, we can then write the structure functions $g_1^p$, $g_1^n$ 
appearing in DIS off polarized proton and neutron targets as
\begin{eqnarray}
2 g_1^p (x,Q^2) &=& \frac{4}{9} \Delta {\cal U} (x,Q^2)+ \frac{1}{9} 
\left[ \Delta {\cal D} (x,Q^2)+ \Delta {\cal S} (x,Q^2)\right] \nonumber \\
2 g_1^n (x,Q^2) &=& \frac{4}{9} \Delta {\cal D} (x,Q^2)+ \frac{1}{9} 
\left[ \Delta {\cal U}(x,Q^2) + \Delta {\cal S} (x,Q^2)\right]  \; ,
\end{eqnarray}
where all parton densities refer to the proton. We can compactly rewrite
this as 
\begin{equation}
g_1^{p,n} (x,Q^2) = \pm \frac{1}{12} \Delta {\cal A}_3  (x,Q^2) + 
\frac{1}{36} \Delta {\cal A}_8  (x,Q^2) + \frac{1}{9} \Delta \Sigma 
(x,Q^2)  \, ,
\end{equation}
where the upper sign refers to the proton, and where we have introduced the
flavor--non-singlet combinations 
$\Delta {\cal A}_3 =\Delta {\cal U}-\Delta {\cal D}$,
$\Delta {\cal A}_8 =\Delta {\cal U}+\Delta {\cal D}-2 \Delta {\cal S}$,
and the singlet $ \Delta \Sigma =\Delta {\cal U}+\Delta {\cal D}+
\Delta {\cal S}$. Had we data at only one $Q^2$, the two structure 
functions $g_1^{p,n}$ could not provide enough information to determine
the full set $\Delta {\cal A}_3,\Delta {\cal A}_8,\Delta \Sigma$ 
at this $Q^2$. When information at
different $Q^2$ is available, one can combine the data with knowledge
about QCD evolution. In particular, each non-singlet quantity evolves 
separately from all other quantities, whereas $\Delta \Sigma$ mixes with
the polarized gluon density $\Delta g(x,Q^2)$ in terms of a matrix
evolution equation~\cite{altarelli77,ahmed76}. Thanks to this property under 
evolution, $g_1^{p,n} (x,Q^2)$ give in principle access to all four 
quantities, $\Delta {\cal A}_3,\Delta {\cal A}_8,\Delta \Sigma,$ and
$\Delta g$~\cite{altarelli97,leader98}. We note that, when 
performing fits to data in practice, one usually also includes constraints 
on the ``first moments'' (Bjorken-$x$ integrals) of $\Delta {\cal A}_{3,8}$ 
derived from the $\beta$-decays of the baryon octet, the 
constraint on 
$\Delta {\cal A}_3$ being essentially the Bjorken sum rule~\cite{bjorken66}.
In this way, one is also able to better determine the first moment of 
$\Delta \Sigma$, which corresponds to the fraction of the proton spin 
carried by quarks and antiquarks.

Information on $\Delta {\cal A}_3,\Delta {\cal A}_8,\Delta \Sigma,$ and $\Delta g$
is equivalent in a ``three-flavor world'' to information on 
$\Delta {\cal U},\Delta {\cal D},\Delta {\cal S},$ and $\Delta g$---this is what
DIS data can provide in principle. We emphasize again that
inclusive DIS cannot give information on the quark and antiquark densities 
separately; it always determines only the $\Delta {\cal Q}$. To distinguish 
quarks from antiquarks, let alone to achieve a full flavor separation of the 
polarized sea, one needs to defer to other processes (see Section~\ref{wasy}).

\begin{figure}[h]
\begin{center} 
   \psfig{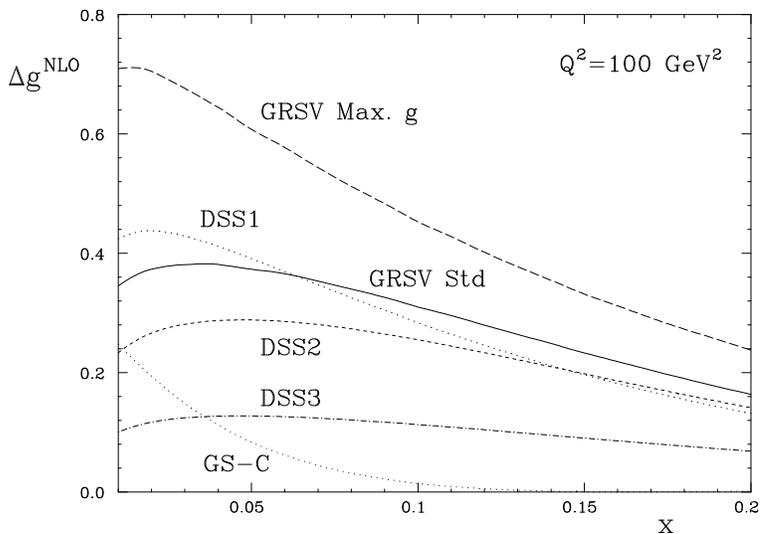}
\end{center}
\caption{
   The polarized gluon densities as
   given by six different NLO 
parameterizations~\protect\cite{gluck96,gehrmann96,deflorian98} ,
   at the scale $Q = 10\ \mathrm{GeV}$. }
\label{figPDF}
\end{figure}

We do not address in detail the question of how well the present data,
within their accuracies, do indeed constrain the quantities $\Delta {\cal A}_3,
\Delta {\cal A}_8,\Delta \Sigma,$ and $\Delta g$. For this we refer the reader to 
the growing number of phenomenological analyses of the polarized 
DIS data~\cite{gluck96,gehrmann96,altarelli97,deflorian98,adams97,leader98}. 
However, to give a very rough picture
of the situation, we state that ($a$) $\Delta {\cal A}_3(x,Q^2)$ and 
$\Delta \Sigma(x,Q^2)$ are relatively well known in the kinematic regions 
where data exist; ($b$) the Bjorken sum rule~\cite{bjorken66} 
is confirmed by the 
data; ($c$) the first moment of $\Delta \Sigma$, and thus the 
quark-plus-antiquark spin contribution to the proton spin, 
is of the order of  25\% or less (known as ``spin surprise''); and ($d$) 
$\Delta {\cal A}_8(x,Q^2)$ and the spin gluon density $\Delta g(x,Q^2)$ 
are 
constrained very little by the data so far. Note that this finding 
for $\Delta {\cal A}_8$ implies also that the polarized strange density is 
still unknown to a large extent. The present situation concerning 
$\Delta g$ is represented by Figure~\ref{figPDF}, which compares the polarized 
gluon densities of several recent NLO sets of spin-dependent parton
distributions~\cite{gluck96,gehrmann96,deflorian98}, all consistent with
current DIS data.
The wide range of possible gluon polarization 
expressed by the figure does not come as a surprise. For DIS,
the gluon is only determined through the scaling violations 
of the structure functions $g_1^{p,n}$; however, so far only 
fixed-target polarized DIS experiments have been carried out, 
which have a limited lever arm in $Q^2$. 
The measurement of $\Delta g$ remains one of the most interesting 
challenges for future high-energy experiments with polarized nucleons.

\def\np#1#2#3{{\it Nucl.\ Phys. }#1:#2 (19#3)}
\def\nc#1#2#3{{\it Nuovo Cim. } #1:#2 (19#3)}
\def\pl#1#2#3{{\it Phys.\ Lett. } #1:#2 (19#3)}
\def\pr#1#2#3{{\it Phys.\ Rev.\ D} #1:#2 (19#3)}
\def\prl#1#2#3{{\it Phys.\ Rev.\ Lett. }#1:#2 (19#3)}
\def\prep#1#2#3{{\it Phys.\ Rep. }#1:#2 (19#3)}
\def\zp#1#2#3{{\it Z.\ Phys.\ C}#1:#2 (19#3)}
\def\rmp#1#2#3{{\it Rev.\ Mod.\ Phys.} #1:#2 (19#3)}
\def\hepph#1{hep-ph/#1}
\def\epj#1#2#3{{\it Eur.\ Phys.\ J.\ C} #1:#2 (19#3)}
\def\etal{et~al}

%
\end{document}